\documentclass[11pt,a4paper]{article}

\usepackage[T1]{fontenc} 
\usepackage{amsmath}
\usepackage{amsfonts}
\usepackage{amssymb}
\usepackage{graphicx}
\usepackage{hyperref}
\usepackage[font=footnotesize,labelfont=bf]{caption}
\begin{document}
\def\thefootnote{\fnsymbol{footnote}}

\begin{center}
\Large{\textbf{Late Time Acceleration of the 3-Space in a Higher Dimensional Steady State Universe in Dilaton Gravity}} \\[0.5cm]
 
\large{\"{O}zg\"{u}r Akarsu$^{\rm a,\,b}$, Tekin Dereli$^{\rm a}$}
\\[0.5cm]

\small{
\textit{$^{\rm a}$ Department of Physics, Ko\c{c} University, 34450 Sar{\i}yer, {\.I}stanbul, Turkey}}

\vspace{.2cm}

\small{
\textit{$^{\rm b}$ Abdus Salam International Centre for Theoretical Physics, Strada Costiera 11, 34151, Trieste, Italy}}

\end{center}

\vspace{.8cm}

\hrule \vspace{0.3cm}
\noindent \small{\textbf{Abstract}\\ 
We present cosmological solutions for $(1+3+n)$-dimensional steady state universe in dilaton gravity with an arbitrary dilaton coupling constant $w$ and exponential dilaton self-interaction potentials in the string frame. We focus particularly on the class in which the 3-space expands with a time varying deceleration parameter. We discuss the number of the internal dimensions and the value of the dilaton coupling constant to determine the cases that are consistent with the observed universe and the primordial nucleosynthesis. The 3-space starts with a decelerated expansion rate and evolves into accelerated expansion phase subject to the values of $w$ and $n$, but ends with a Big Rip in all cases. We discuss the cosmological evolution in further detail for the cases $w=1$ and $w=\frac{1}{2}$ that permit exact solutions. We also comment on how the universe would be conceived by an observer in four dimensions who is unaware of the internal dimensions and thinks that the conventional general relativity is valid at cosmological scales.}
\\
\noindent
\hrule
\noindent \small{\\
\textbf{Keywords:} Kaluza-Klein cosmology $\cdot$ Higher dimensional dilaton gravity $\cdot$ Late time acceleration}
\def\thefootnote{\arabic{footnote}}
\setcounter{footnote}{0}
\let\thefootnote\relax\footnote{\textbf{E-Mail:} oakarsu@ku.edu.tr,  tdereli@ku.edu.tr}

\def\thefootnote{\arabic{footnote}}
\setcounter{footnote}{0}

\section{Introduction}
\label{intro}

The accelerated expansion of the universe first came under scrutiny
in 1917 with de Sitter's vacuum solution of the Einstein's field
equations with a positive cosmological constant. However, it has not
been considered seriously before the inflationary
scenario \cite{Guth} that is characterized by an epoch of exponential
expansion in the very early universe at energy scales $\sim
10^{16}\,{\rm GeV}$ was proposed in 1980. The inflation idea is one
of the most prominent attempts to resolve the problems of standard
Big Bang Cosmology such as the observed spatial homogeneity,
isotropy, and flatness of the universe and others (See \cite{Linde08}
for a recent review).  The idea has many variations which are
generally based on conventional general relativity theory together
with a number of scalar fields introduced in an \textit{ad hoc} way.
Yet it is still a scenario as a concrete derivation of inflation
from a fundamental theory such as the string theory \cite{Quevedo02}
hasn't been found.  On the other hand, the fact that our universe has entered a state of accelerated expansion at redshifts less than $\sim 0.5$ is well established by independent studies \cite{Riess98,Perlmutter99,Alam04,Sahni06,Tasos09,Lima10,Cunha09,Li11,Komatsu11}. The lack of a
satisfactory physical explanation of the current acceleration is a
mystery because it happens at energy scales $\sim 10^{-4}\,{\rm eV}$
where we supposedly  know physics very well. The observed dynamics
of the universe could be accommodated surprisingly well in
$\Lambda$CDM cosmology by the inclusion of a positive cosmological
constant into the Einstein's field equations in the presence of
pressure-less matter. But now the cosmological constant problem has
to be faced  which is one of the pressing unsolved problems in
physics today \cite{Zeldovich,Weinberg89,Sahni00}: The observed
energy density of the vacuum turns out to be non-zero but some 120
orders of magnitude less than the theoretical estimates. This is
about the same order of magnitude as of the baryonic matter, a fact
that makes the present epoch of the universe very special
(coincidence problem).

The efforts to solve cosmological problems in the context of
conventional general relativity led many to consider various sources
of dark energy. However, the known dark energy candidates such as
quintessence, k-essence, phantom, quintom and others are mostly ad
hoc and/or considered phenomenologically rather than being derived
from a fundamental theory \cite{Copeland06,Bamba12}. Another approach to cope
with the current acceleration of the universe is to consider the
possibility that the Einstein's theory of gravity (viz. general
relativity) is not an adequate theory for understanding gravitation
at cosmological scales. There are many alternative  theories of
gravity suggested with extra fields (scalar-tensor, Einstein-aether,
...), including higher derivatives and non-local terms in the action
( $R^2$, $f(R)$, ...) or living in higher dimensions (Kaluza-Klein,
Randall-Sundrum, ...) (See \cite{Capozziello11extended,Nojiri11,Clifton12}
for comprehensive reviews). In fact any successful unified theory of
everything, such as a superstring theory, is expected to resolve all
the unexplained issues  once and for all. Moreover, cosmology may
turn out to be the main avenue to probe string theories, unless
supersymmetry is discovered at low energies and its properties
provide some hints to its high energy origin or else the string
scale turns out to be low enough to be probed at LHC or some future
collider \cite{Quevedo02}.

Motivated by the arguments above, we consider here a higher
dimensional dilaton gravity theory that may arise as the low energy
effective field theory approximation to a superstring theory.
It is well known that bosonic string theories require a
26-dimensional space-time for consistency while  anomaly-free
superstring theories live in 10 dimensions. It is generally assumed
in the higher dimensional context that all but four of the
space-time dimensions are compactified on an unobservable internal
manifold,  leaving an observable effective action to be considered
in a $(1+3)$-dimensional space-time \cite{Copeland,Lidsey}. There is
yet another approach to the problem of extra dimensions. They may be
relaxed to become dynamical so that they start at larger scales in
the early universe and evolve in time to unobservable scales, thus
leading to an effective four dimensional picture of the universe
(See \cite{OverduinWesson97} for a comprehensive review on
Kaluza-Klein cosmologies). This approach has been considered for the
first time in the early 1980s as a dynamical reduction of internal
dimensions with the external dimensions expanding while the internal
dimensions
contracting \cite{ChodosDetweiler80,Freund82,DereliTucker83}. Among
the higher dimensional cosmological models, there is an interesting
subclass that describe higher dimensional steady state cosmologies
with the total volume of the higher dimensional space remaining
constant in time \cite{Freund82,DereliTucker83,BleyerZhuk96,RainerZhuk00,HoKephart10}.
In these papers, the
external space expands with a constant deceleration parameter which
implies either a power-law expansion or exponential expansion. In a
recent study \cite{AkarsuDereli12HDSS}, we have assumed from the beginning that both
the higher dimensional volume of the universe and the  energy
density are constants and found some new solutions where the
external space exhibits dynamical behavior with time dependent
deceleration. We further concluded that the model may be improved by considering modified gravity theories. Indeed, in this paper considering a higher dimensional dilaton gravity we not only present power-law dynamics but also dynamics in which the external space accommodates the observed universe starting from the primordial nucleosynthesis epoch to the present universe. In all the models that can accommodate the observed universe, on the other hand, the external space ends with a Big Rip and hence lives for a finite time. This is also a result consistent with the observational studies that argue for the Big Rip behavior in the future of the universe is slightly favored when compared, for instance, with the de Sitter future of the $\Lambda$CDM cosmology \cite{Godlowski05,Zhao07,Alberto12,Novosyadlyj12,Parkinson12}. Moreover, because the universe would live for a finite time, the coincidence problem will not be a severe problem anymore, as discussed first by Caldwell et al. \cite{Caldwell03}.

The organization of the paper is as follows. In Section \ref{Fieldequations}, we
introduce the dilaton gravity action in $(1+3+n)$-dimensions written
in the string frame. The variations of the action and the
cosmological field equations are also given. We discuss in Section \ref{ConstantVol}
steady state cosmological models with a constant $(3+n)$-dimensional
volume. We give two classes of exact solutions with the external
space expanding while the internal space is contracting in Section \ref{sec:w1}
and Section \ref{sec:w1b2} and study their observational consequences separately
in detail. A brief discussion on how the universe appears to an
observer living in four dimensions is given in Section \ref{ObserverUniverse}. We conclude
by some final comments.

%%%%%%%%%%%%%%%%%%%%%%%%%%%%%%%%%%%%%%%%%%%%%%%%%%%%%%%%%%%%%%%%%%%

\section{Dilaton Gravity Equations}
\label{Fieldequations}

Inspired by the low-energy effective string theory, we consider the following dilaton gravity action in $(1+3+n)$-dimensions written in the {\sl string frame}:
\begin{equation}
\label{eqn:action}
I = \int_{\mathcal{M}} {\rm d}^{1+3+n}x \sqrt{|g|} e^{-2 \varphi} \left \{ \mathcal{R}+ 4 w \; \partial_{\mu} \varphi \partial^{\mu} \varphi + U(\varphi)\right \}.
\end{equation}
$\mathcal{R}$ is the scalar curvature of the space-time metric $g$ and $\varphi$ is the dilaton field. The dilaton self-interaction potential $U(\varphi)$ is a real, smooth function of the dilaton to be specified. We introduce an arbitrary positive dilaton coupling constant $w>0$ to cover a variety of cases in the string frame. For instance choosing $w=1$, we obtain gravi-dilaton string effective action with a dilaton self-interaction potential in $(1+3+n)$-dimensions and the further choices $1+3+n=10$ and $1+3+n=26$ correspond to anomaly-free superstring and bosonic string theories, respectively. Our model in the string frame will certainly differ from similar models in the Jordan or Einstein frames. However, since the string frame is the natural frame with respect to string models and since there is no a priori argument that would favor one frame over the others, here we seek a fit to the observed universe in the string frame \cite{Easson99}. We should emphasize that we do not introduce any higher-dimensional matter source in the action (\ref{eqn:action}). This is in line with the very central idea of string models where matter in four-dimensions is nothing but a manifestation of strings that live in higher dimensions.

We consider a spatially homogeneous but not necessarily isotropic $(1+3+n)$-dimensional synchronous space-time metric that involves a maximally symmetric three dimensional flat external space metric and a compact $n$-dimensional flat internal space metric:
\begin{equation}
\label{eqn:HDmetric}
{\rm d}s^2 = -{\rm d}t^2 + a^2(t) \left ( {\rm d}x^2 + {\rm d}y^2 +{\rm d}z^2 \right ) + s^2(t)
\left ( {\rm d}\theta_{1}^{2} +...+  {\rm d}\theta_{n}^{2}\right ).
\end{equation}
Here $t$ is the cosmic time variable, $(x,y,z)$ are the Cartesian coordinates of the 3-dimensional flat space that represents the space we observe today, and $(\theta_1, ...,\theta_n)$ are the coordinates of the $n$-dimensional, compact (toroidal) internal space that represents the space that cannot be observed directly and locally today. The scale factors $a(t)$ of the 3-dimensional external space and $s(t)$ of the $n$-dimensional internal space are functions of $t$ only.

Here we will be considering an exponential potential for the dilaton field
\begin{equation}
U(\varphi) = U_0 e^{\lambda \varphi},
\end{equation}
where $U_0$ and $\lambda$ are two real parameters and the dilaton field is space independent real function of time $t$:
\begin{equation}
\varphi = \varphi(t).
\end{equation}

The field equations to be solved are determined from the action (\ref{eqn:action}) by a variational principle. The usual way to proceed at this stage is to substitute the cosmological field expressions into these field equations and reduce them to a system of coupled ordinary differential equations. Here we will proceed in the following way to reach this final set of equations to be solved. Let us first substitute the cosmological fields into the action integral so that the reduced action reads
\begin{equation}
I[a, s, \varphi]  = \int {\rm d}t \int {\rm d}^3x \int {\rm d}^n\theta \mathcal{L}
\end{equation}
where the Lagrangian function
\begin{equation}
\mathcal{L} = e^{-2\varphi} a^3 s^n \left \{\mathcal{R} - 4w
{{\varphi}^{\prime}}^2 + U_0 e^{\lambda \varphi} \right \}
\end{equation}
with the curvature scalar given by
\begin{equation}
\nonumber
 \mathcal{R} = 6 \left ( \frac{a^{\prime }}{a}\right )^2+ 6 \frac{a^{\prime \prime}}{a} 
+ n(n-1) \left (\frac{s^{\prime }}{s}\right )^2 + 2n \frac{s^{\prime
 \prime}}{s}+ 6n \frac{a^{\prime }}{a} \frac{s^{\prime }}{s}.
\end{equation}
Here ${}^{\prime}$ denotes $\frac{\rm d}{{\rm d}t}$. Differentiating by parts and neglecting total derivatives we arrive at an equivalent Lagrangian function
\begin{equation}
\tilde{\mathcal{L}} = e^{-2\varphi} a^3 s^n \left \{ -6n
\frac{a^{\prime }}{a} \frac{s^{\prime }}{s} - 6 \left (
\frac{a^{\prime }}{a}\right )^2  -n(n-1) \left ( \frac{s^{\prime }}{s}
\right )^2 + 12 \frac{a^{\prime }}{a} \varphi^{\prime } +4n
\frac{s^{\prime }}{s} \varphi^{\prime }  - 4w {{\varphi}^{\prime}}^2
+ U_0 e^{\lambda \varphi} \right \}
\end{equation}
from which we get the following  Euler-Lagrange equations of motion
\begin{eqnarray}
\label{eqn:12}
3\frac{{a'}^2}{a^2}+3\frac{a''}{a}+(3n-3)\frac{{a'}}{a}\frac{{s'}}{s}+\left(\frac{1}{2}n^2-\frac{3}{2}n+1\right)\frac{{s'}^2}{s^2}+(n-1)\frac{s''}{s}-2{\varphi}''\\
\nonumber
+(4-2w){{\varphi}'}^2
-2{\varphi}'\left(3\frac{a'}{a}
+(n-1)\frac{{s'}}{s}\right)
+\frac{1}{2} U_{0} e^{\lambda \varphi}
&=0,
\end{eqnarray}
\begin{equation}
\label{eqn:13}
3\frac{{a'}^2}{a^2}+3n\frac{{a'}}{a}\frac{{s'}}{s}+\frac{n(n-1)}{2}\frac{{s'}^2}{s^2}+3\frac{a''}{a}+n\frac{s''}{s}-2w{\varphi}''+2w{{\varphi}'}^2-2w{\varphi}'
\left(3\frac{a'}{a}+n\frac{s'}{s}\right)+
\frac{2-\lambda}{4}U_{0}e^{\lambda \varphi}=0,
\end{equation}
\begin{equation}
\label{eqn:14}
\frac{{a'}^2}{a^2}+2\frac{a''}{a}+2n\frac{{a'}}{a}\frac{{s'}}{s}+n\frac{s''}{s}+\frac{n(n-1)}{2}\frac{{s'}^2}{s^2}-2{\varphi}''+(4-2w){{\varphi}'}^2-2{\varphi}'
\left(2\frac{a'}{a}+n\frac{s'}{s} \right)+\frac{1}{2}
U_{0}e^{\lambda\varphi}=0,
\end{equation}
for the dynamical variables $a$, $\varphi$ and $s$,  respectively. These coupled equations must be supplemented by the Hamiltonian constraint equation
\begin{equation}
\label{eqn:16}
3\frac{{a'}^2}{a^2}+3n\frac{{a'}}{a}\frac{{s'}}{s}+\frac{n(n-1)}{2}\frac{{s'}^2}{s^2}+2w{{\varphi}'}^2-2{\varphi}'
\left(3\frac{a'}{a}+n\frac{s'}{s}\right)+\frac{1}{2}U_{0}
e^{\lambda\varphi} = 0 .
\end{equation}
We would like to note that any one of the equations out of (\ref{eqn:12})-(\ref{eqn:16}) can be derived from the remaining two other equations plus the Hamiltonian constraint equation (See, for instance, \cite{Dereli93} for details).

\bigskip

\section{Cosmological Models with $(3+n)$-Dimensional Constant Volume}
\label{ConstantVol}

We are interested in the solutions for a higher dimensional steady state universe. In a recent study \cite{AkarsuDereli12HDSS}, we define a higher dimensional steady-state universe with two properties: (i) the higher dimensional universe has a constant volume as a whole but the internal and external spaces are dynamical, (ii) the energy density is constant in the higher dimensional universe. However in this study, we do not introduce any matter source in higher dimensions and hence it is enough if we consider only the first one of these properties. Accordingly, we assume that the volume scale factor of the higher dimensional universe, i.e. the $(3+n)$-dimensional volume, is constant:
\begin{equation}
\label{eqn:constvol}
V=a^{3}s^{n}=V_{0}.
\end{equation}
$V_{0}$ is a positive real constant. We note that this condition assures the dynamical contraction (hence reduction) of the internal space for an expanding external space and that the constant $(3+n)$-dimensional volume may be interpreted either as the total volume of the $(3+n)$-dimensional space of the $(1+3+n)$-dimensional universe or as the volume of a $(3+n)$-dimensional space-like domain in a $(1+3+n)$ dimensional universe. Substituting $(3+n)$-dimensional constant volume constraint \eqref{eqn:constvol} in the equations \eqref{eqn:12}-\eqref{eqn:16}, they reduce to the following set of equations:
\begin{equation}
\label{eqn:12b}
\frac{3}{2}\left(\frac{1}{n}+1\right)\frac{{a'}^2}{a^2}+\frac{3}{n}\frac{a''}{a}-2{\varphi}''+(4-2w){{\varphi}'}^2-\frac{6}{n}{{\varphi}'}\frac{a'}{a}+\frac{1}{2} U_{0} e^{\lambda \varphi}=0,
\end{equation}
\begin{equation}
\label{eqn:13b}
\frac{3}{2}\left(\frac{3}{n}+1\right)\frac{{a'}^{2}}{a^2}-2w{{\varphi}''} +2w{{\varphi}'}^{2}+\frac{2-\lambda}{4}U_{0} e^{\lambda \varphi} = 0,
\end{equation}
\begin{equation}
\label{eqn:14b}
\left(\frac{9}{2n}+\frac{5}{2}\right)\frac{{a'}^{2}}{a^2}-\frac{{a}''}{a}-2{{\varphi}''}+(4-2w){{\varphi}'}^{2}+2\frac{{a}'}{a}{{\varphi}'}+\frac{1}{2}U_{0}e^{\lambda \varphi}=0,
\end{equation}
\begin{equation}
\label{eqn:16b}
-\frac{3}{2}\left(\frac{3}{n}+1\right)\frac{{a'}^{2}}{a^2}+2w{{\varphi}'}^{2}+\frac{1}{2}U_{0}e^{\lambda \varphi} = 0.
\end{equation}
Now adding equations \eqref{eqn:13b} and \eqref{eqn:16b} side by side, the dilaton field decouples from the scale factors and satisfies 
\begin{equation}
\label{eqn:18}
16w{{\varphi}'}^2+\, (4-\lambda)U_{0} e^{\lambda\varphi}-8w{\varphi}''=0.
\end{equation}
Furthermore subtracting \eqref{eqn:14b} from \eqref{eqn:12b} and then using \eqref{eqn:18} we obtain
\begin{equation}
{\varphi}'=\frac{1}{2}\frac{a''}{a'}-\frac{1}{2}\frac{a'}{a};
\end{equation}
which finally gives
\begin{equation}
\label{eqn:apure}
a=a_{0}e^{c_{1}\int{e^{2\varphi}}\mathrm{d}t}
\end{equation}
where $a_{0}$ and $c_{1}$ are real integration constants. Now subtracting \eqref{eqn:13b} from \eqref{eqn:14b} and substituting \eqref{eqn:apure} we obtain
\begin{equation}
c_{1} n (\lambda-4+4w)\left(2{{\varphi}'}^{2}-{\varphi}''\right)=0.
\end{equation}
We are looking for solutions where the universe is possibly dynamical. Then $c_{1}\neq 0$ from (\ref{eqn:apure}) and we must set $n\neq0$, otherwise there will no internal dimensions. Hence a dynamical cosmological model with constant $(3+n)$-dimensional volume scale factor constraint is not consistent for arbitrary values of the constants $w$ and $\lambda$, but is subject to the condition
\begin{equation}
\label{eqn:wlambda}
\lambda=4-4w.
\end{equation}
This condition reduces \eqref{eqn:18} to the following second-order non-linear ordinary differential equation
\begin{equation}
\label{eqn:varphifinal}
{\varphi}''-2{{\varphi}'}^2-\frac{\, U_{0}}{2} e^{4(1-w)\varphi}=0.
\end{equation}
Now defining a new scalar field
\begin{equation}
\label{eqn:beta}
\beta (t)=e^{-2\varphi},
\end{equation}
which is a positive function since $\varphi$ is a real function, and using this in (\ref{eqn:varphifinal}) we obtain the following second-order autonomous differential equation
\begin{equation}
\label{eqn:betadiffeqn}
\beta''=-U_{0}\beta^{2w-1}.
\end{equation}
Integrating this equation once we obtain
\begin{equation}
\label{eqn:firstintbeta}
2\mathcal{E}={\beta'}^2+\frac{U_{0}}{w}\beta^{2w}
\end{equation}
where $\mathcal{E}$ is a real constant of integration. This differential equation is familiar from the classical central force problem in one dimension in the presence of a power-law potential, though with a slight difference. One may observe from (\ref{eqn:beta}) that $\beta$ is non-negative $\beta\geq 0$, since dilaton field $\varphi$ is defined to be a real function. This should be kept in mind in what follows as we proceed with the solutions considering $\beta\geq 0$. We would like to expose the meaning of the integration constant $\mathcal{E}$. Equation (\ref{eqn:firstintbeta}) can be rewritten as
\begin{equation}
\mathcal{E}=\frac{1}{2}{\beta'}^2-\int {-U_{0}\beta^{2w-1}{\rm d}\beta}.
\end{equation}
It is of the form
\begin{equation}
\label{eqn:k1}
\mathcal{E}=\mathcal{T}-\int {\mathcal{F}{(\beta)}{\rm d}\beta}=\mathcal{T}+\mathcal{U}(\beta),
\end{equation}
where $\mathcal{T}$ is the kinetic energy and $\mathcal{U}$ is the potential energy given by
\begin{equation}
\mathcal{T}=\frac{1}{2}{\beta'}^2,\quad
\mathcal{U}(\beta)=-\int{\mathcal{F}(\beta)}{\rm d}\beta=\frac{U_{0}}{2w}\beta^{2w},
\end{equation}
where $\mathcal{F}(\beta)=-U_{0}\beta^{2w-1}$. Hence, $\mathcal{E}$ corresponds to the total energy of the $\beta$ field. From (\ref{eqn:firstintbeta}) we have
\begin{equation}
\label{eqn:betaint}
\int{\frac{{\rm d}\beta}{\sqrt{2\mathcal{E}-\frac{U_{0}}{w}\beta^{2w}}}}=\pm t-t_{0},
\end{equation}
where $t_{0}$ is an integration constant and can be chosen arbitrarily since it is nothing but a shift in the cosmic time $t$. For convenience we choose $t_{0}=0$ and positive sign for the time parameter $t$.

Finally, a relation between the scale factor of the external space $a$ and the $\beta$ field (hence dilaton field $\varphi$), including the dilaton coupling constant $w$, the total energy $\mathcal{E}$ and the number of the internal dimensions $n$ can now be given:
\begin{equation}
\label{eqn:16final}
\frac{{a'}^{2}}{a^2} =\frac{n}{3(3+n)}  \frac{2w \mathcal{E}}{\beta^2}.
\end{equation}
We note that $\mathcal{E}\geq 0$, since $w>0$.  Integrating \eqref{eqn:16final} we obtain the scale factor of the external space, and hence of the internal space via (\ref{eqn:constvol}), as follows
\begin{equation}
\label{eqn:scales}
a=a_{0} e^{\pm\frac{1}{3}\sqrt{\frac{3n}{(3+n)}} \sqrt{2w \mathcal{E}}  \int{\frac{1}{\beta} {\rm d}t}}
\quad\textnormal{and}\quad
s=s_{0} e^{\mp\frac{1}{n}\sqrt{\frac{3n}{(3+n)}} \sqrt{2w \mathcal{E}}  \int{\frac{1}{\beta} {\rm d}t}}.
\end{equation}
$a_{0}$ and $s_{0}$ are integration constants that satisfy constant volume condition $V_{0}=a_{0}^{3} s_{0}^n$. Hence, the scale factors may be obtained explicitly whenever $\beta$ can be obtained explicitly from (\ref{eqn:betaint}). However, in general it is not possible to evaluate the integral in (\ref{eqn:betaint}) in terms of known functions for arbitrary values of $w$, $U_{0}$ and $\mathcal{E}$, but only for particular choices. It is possible to evaluate the integrals analytically and isolate $\beta$ for arbitrary values of $w$ in two cases: (i) when the total energy of the $\beta$ field is zero $\mathcal{E}=0$ and the potential of the dilaton field is non-zero $U_{0}\neq 0$, (ii) when the dilaton field has no potential $U_{0}=0$ and the total energy of the $\beta$ field is positive $\mathcal{E}> 0$. On the other hand, when $\mathcal{E}> 0$ and $U_{0}\neq 0$, it is possible to evaluate the integral and isolate $\beta$ for $w=1,\; \frac{1}{2}$; for other values of $w>0$ it is possible to obtain results for $\beta$ in terms of circular or elliptic functions but in many cases only numerical results maybe gained. Fortunately, we are able to get some information for arbitrary values of $w>0$ that may be important from the cosmological point of view. To do so, we first give some important cosmological parameters in the terms of $\beta$ and its derivatives with respect to time. Using (\ref{eqn:scales}) we obtain the Hubble parameters of the external and internal spaces, respectively, as follows:
\begin{equation}
\label{eqn:Hbeta}
H_{a}\equiv\frac{a'}{a}=\pm\frac{1}{3}\sqrt{\frac{3n}{3+n}} \frac{\sqrt{2w \mathcal{E}}}{\beta}
\quad\textnormal{and}\quad
H_{s}\equiv\frac{s'}{s}=\mp\frac{1}{n}\sqrt{\frac{3n}{3+n}} \frac{\sqrt{2w \mathcal{E}}}{\beta}.
\end{equation}
We note that Hubble parameters are inversely proportional with $\beta$ and either the external space expands while the internal space contracts or conversely the external space contracts while the internal space expands such that $(3+n)$-dimensional volume scale factor is constant, i.e., $\dot{V_{0}}/V_{0}=3H_{a}+nH_{s}=0$.  Because we are interested in the solutions where the external space expands in particular in what follows we shall dwell on solutions with expanding external space, i.e., with the cases $H_{a}>0$. The dimensionless deceleration parameter is another important cosmological parameter and the deceleration parameters of the external and of the internal dimensions are
\begin{equation}
\label{eqn:dpbeta}
q_{a}\equiv -\frac{a'' a}{{a'}^{2}}= -1\pm3\sqrt{\frac{3+n}{3n}}\frac{\beta'}{\sqrt{2w\mathcal{E}}}
\quad\textnormal{and}\quad
q_{s}\equiv -\frac{s'' s}{{s'}^{2}}=-1\mp n\sqrt{\frac{3+n}{3n}}\frac{\beta'}{\sqrt{2w\mathcal{E}}},
\end{equation}
respectively. The deceleration parameter is particularly important when we are discussing the dynamics of the external space that represents the space we observe today. The negative values of deceleration parameter correspond to the acceleration and positive values to the deceleration of the scale factor. The space is said to exhibit super-exponential expansion for $q<-1$; exponential expansion (also known as the de Sitter expansion) for $q=-1$;  accelerated power-law expansion for $-1<q<0$; expansion at a constant rate for $q=1$ and finally decelerated expansion for  $q>1$. In the $\Lambda$CDM cosmology, which is the simplest model that accommodates the observed dynamics of the universe but with some serious problems, the deceleration parameter evolves from $\frac{1}{2}$ to $-1$ as the universe expands  \cite{Sahni00}. In most of the scalar field dark energy models in the context of $(1+3)$-dimensional general relativity, on the other hand, it evolves between $2$ and $-1$ \cite{Copeland06}. Because we are concerned in this paper with an evolving deceleration parameter, it is convenient to discuss also the dimensionless jerk/state finder parameter (that involves the third derivative of the scale factor of  the observed universe with respect to time) of the external space:
\begin{equation}
j_{a}\equiv\frac{a''' a^2}{{a'}^{3}}=\pm 6\frac{3+n}{n} \left(2\beta'-9\sqrt{\frac{3+n}{3n}} \beta'-\beta'' \beta+1 \right) w \mathcal{E}.
\end{equation}
The jerk parameter in $\Lambda$CDM is simply equal to unity, $j_{\Lambda{\rm CDM}}=1$ \cite{Sahni03}. Hence the search for variation of $j$ from unity over some redshift interval or time interval is also very convenient for seeing the departures from $\Lambda$CDM model.
However, the observational constraints on the value of jerk parameter are rather weak in the present literature. The various observational studies give the observed value of the jerk parameter approximately in the range $-5\lesssim j_{a,{\rm today}}\lesssim 10$ \cite{Sahni03,Visser04,Rapetti07,Cattoen08,Wang09,Vitagliano10,Capozziello11,Xia12}.

\subsection{The Static Universe}
\label{static}
The first case we would like to discuss is the one where $\mathcal{E}=0$ and $U_{0}\neq 0$, i.e. the total energy density of the $\beta$ field is null $\mathcal{E}=0$ but the potential is non-zero $U_{0}\neq 0$. However, we note from (\ref{eqn:firstintbeta}) that the case $\mathcal{E}=0$ is possible only if $U_{0}<0$. Now choosing $\mathcal{E}=0$ in (\ref{eqn:betaint}) we obtain:
\begin{equation}
\beta=\left[-\frac{(1-w)^{2}U_{0}}{w}\;t^2\right]^{\frac{1}{2(1-w)}}
\quad\textnormal{and}\quad
\varphi=-\frac{1}{4(1-w)}\ln\left(-\frac{(1-w)^{2}U_{0}}{w}\;t^2\right)
\quad\text{for}\quad w > 1
\end{equation}
and
\begin{equation}
\beta=e^{ \sqrt{-U_{0}}\;t}
\quad\textnormal{and}\quad
\varphi= \frac{\sqrt{-U_{0}}}{2}t
\quad\text{for}\quad w= 1.
\end{equation}
On the other hand, although the dilaton field is dynamical, the universe is non-dynamical in this case. Choosing $\mathcal{E}=0$ in (\ref{eqn:scales}) we have
\begin{equation}
a=a_{0}\quad\textnormal{and}\quad s=s_{0}.
\end{equation}
This result tells us that if $\mathcal{E}=0$, that is, if the total energy of the $\beta$ field is zero, then we have a static universe independent of the value of dilaton coupling constant $w$.

\subsection{Dynamical universes}
\label{Dynamical}
Before giving the solutions with dynamical external and internal spaces we would like to give a short discussion on whether we can have consistent effective 4-dimensional cosmologies or not under the assumption of constant higher dimensional volume independent of a specific solution.

We first comment on the role of the number of internal dimensions $n$ on the expansion rate of the external dimensions and the contraction rate of the internal dimensions and their sizes. From the constant $(3+n)$-dimensional volume condition, the magnitude of the expansion rate of the external space and of the contraction rate of the internal space are identical independent of the number of the internal dimensions. However, one may observe from (\ref{eqn:Hbeta}) that the number of the internal dimensions $n$ appears as two different factors in the Hubble parameters of the external and internal dimensions and affects the expansion rate of the external dimensions slightly but the contraction rate of the internal dimensions drastically so that for very large $n$ values the internal dimensions are almost frozen. Apparently, this has consequences on the size of the external and internal dimensions. From the constant higher dimensional volume condition, the size of the internal dimensions can be given as follows: $s={V_{0}}^{\frac{1}{n}} a^{-\frac{3}{n}}$. This relation gives us opportunity to investigate whether the internal dimensions are involved with the process of primordial nucleosynthesis, in the early universe, that shouldn't be affected by the presence of internal dimensions for any viable cosmological model. To do so, we assume that the size of the internal dimensions remain roughly less than the size of a proton $l_{\rm proton}\sim 1\, {\rm GeV}^{-1} \sim 10^{-16}\,{\rm m}$ when the primordial nucleosynthesis occurs, i.e., $s_{\rm NS}\lesssim l_{\rm proton}$. Additionally we can write
\begin{equation}
\label{eqn:NS1}
s=\left(\frac{a_{\rm today}}{a}\right)^{\frac{3}{n}} s_{\rm today}
\end{equation}
from the constant higher dimensional volume condition (\ref{eqn:constvol}). As long as the wavelength of the cosmic background radiation (CBR) is so high that it cannot propagate into the internal dimensions $s\,{(\rm m)}\lesssim 438\,T^{-1}\,({\rm Kelvin}^{-1}) $ we can use the standard relation\footnote{In fact the relation is as follows $\frac{a_{\rm today}}{a}=\beta\frac{T}{T_{\rm today}}$, where $\beta$ accounts for non-adiabatic expansion due to entropy production. In standard cosmology $\beta=1$ for $T<m_{\rm electron}$ and $\beta=(\frac{11}{4})^{\frac{1}{3}}$ for $T>m_{\rm electron}$. However, we simply use $\beta=1$ in this study, since it is a good approximation within the context of our discussions.  See  Ref. \cite{Kaplinghat99} and references therein for details.} between the CBR temperature and the scale factor of the external space
\begin{equation}
\label{eqn:NS2}
\frac{a_{\rm today}}{a}\sim\frac{T}{T_{\rm today}}.
\end{equation}
Accordingly a successful model should provide the condition
\begin{equation}
\label{eqn:NS3}
s_{\rm NS}=\left(\frac{T_{\rm NS}}{T_{\rm today}}\right)^{\frac{3}{n}} s_{\rm today}\lesssim l_{\rm proton}\sim 10^{-16} {\rm m}.
\end{equation}
The primordial nucleosynthesis of the light elements is determined by the events occurring in the epoch when the temperature of the universe varied from $\sim 1\,{\rm MeV}$ ($\sim 10^{9}\,{\rm K}$) to $\sim 0.1 \,{\rm MeV}$ ($\sim 10^{8}\,{\rm K}$) and the current temperature of the CBR is $T_{\rm today}\cong 2.352\times 10^{-4}$eV ($T_{\rm today}=2.728$ K) \cite{Fixsen09,Komatsu11}. Accordingly, we find that $a_{\rm today}/a_{\rm NS} \sim 10^{9}$ from (\ref{eqn:NS2}) and $s_{\rm today}/s_{\rm NS} \sim 10^{-\frac{27}{n}}$ from (\ref{eqn:NS3}). We first check if the constant higher dimensional volume constraint remains valid under an extremely strong condition such as the case when the size of the internal dimensions today are almost at LHC length scales $s_{\rm today}\sim l_{\rm LHC}\sim 10^{-20} {\rm m}$ ($l_{\rm LHC}\sim 10^{-13} {\rm eV^{-1}}$) and can be observed very soon. In this case the condition $s_{\rm NS}\lesssim l_{\rm proton}$ is satisfied provided $n\gtrsim 7$. On the other hand, assuming the size of the internal dimensions today are at Planck length scale $s_{\rm today}\sim l_{\rm Planck}\sim 10^{-35} {\rm m}$ ($l_{\rm Planck}\sim 10^{-28} {\rm eV^{-1}}$) then $s_{\rm NS}\lesssim l_{\rm proton}$ condition is satisfied provided $n\gtrsim 2$. We can also estimate the mean length scale of the volume of the higher dimensional universe $l=(a^{3}s^{n})^{\frac{1}{3+n}}$ using $s_{\rm NS}\lesssim l_{\rm proton}$ for  $s_{\rm today}\sim l_{\rm LHC}$ and $s_{\rm today}\sim l_{\rm Planck}$. The size of the observed universe today can be taken as $a_{\rm today}\sim 10^{26}\,{\rm m}$. Using $s_{\rm today}\sim l_{\rm LHC}$ and $n\gtrsim 7$ we obtain $l\lesssim 10^{-6}\, {\rm m}$ and using  $s_{\rm today}\sim l_{\rm Planck}$ and $n\gtrsim 2$ we obtain $l\lesssim 10^{0}\, {\rm m}$ for the mean length scale of the higher dimensional constant volume. The analysis above shows that we can have cosmological models with higher dimensional constant volume without the intervention of the internal dimensions to the primordial nucleosynthesis processes. The abundances of light elements we observe today, particularly that of $^4$He, give us the opportunity of probing the dynamics of the early universe. Let us first give a brief summary of Big Bang Nucleosynthesis (BBN) model \cite{Steigman05,Steigman08}. In the early universe when $T\gtrsim 1\,{\rm MeV}$, the charged-current weak interactions among neutrons, protons, electrons, positrons and neutrinos maintain the neutron-proton equilibrium: $\frac{n_{\rm n}}{n_{\rm p}}=e^{-\frac{\Delta m}{T}}$ where $n_{\rm n}$ and $n_{\rm p}$ are the number densities of the neutrons and protons, respectively, and $\Delta m=m_{\rm n}-m_{\rm p}=1.293\,{\rm MeV}$ is the neutron-proton mass difference. Accordingly, there are as many neutrons as protons when $T>> 1\, {\rm MeV}$ and the neutron fraction gets smaller for $T < 1 {\rm MeV}$. If the weak interactions operate indefinitely and efficiently enough to maintain equilibrium, then the neutron abundance would drop to zero. However, if the expansion rate of the universe $H$ and the rate of the weak interactions $\Gamma\sim G_{F}^2 T^5$ ($G_{F}^2$ is the Fermi constant) are of the same order of magnitude $H\sim\Gamma$, then the weak interactions would not operate efficiently anymore and the $\frac{n_{\rm n}}{n_{\rm p}}$ ratio freezes-out. Then the ratio continues to decrease only very slowly, due to out-of-equilibrium weak ineteractions and free neutrons decay with a lifetime of $887\,{\rm s}$. All the while neutrons and protons keep on colliding to form deuterium which are photodissociated by the cosmic background photons (gamma rays at that time). The very low abundance of deuterium removes the platform for building heavier nuclei; thus the nucleosynthesis is delayed by this photodissociation bottleneck. When the temperature drops below $\sim 80\, {\rm keV}$, the deuterium bottleneck would be broken (which should occur before $887\,{\rm s}$) and nuclear reactions quickly burn the remaining free neutrons into $^{4}$He, leaving behind trace amounts of D, $^3$He and $^7$Li. Consequently, virtually all neutrons available will be incorporated into $^4$He whose mass ratio can be given as $Y_{\rm p}\equiv\frac{4n_{\rm ^{4}{\rm He}}}{n_{\rm n}+n_{\rm p}}\approx \frac{2n_{\rm n}}{n_{\rm n}+n_{\rm p}}$. This will be very sensitive to temperature and hence to the expansion rate of the universe at the time of freeze-out. This shows that the value of $Y_{\rm p}$ we observe today gives us information about the expansion rate of the early universe when BBN took place. Therefore, we can examine our models by checking whether they may provide the suitable conditions at early times in accordance with the above model and the predicted $Y_{\rm p}$ values in our models are consistent with the observationally inferred values $Y_{\rm p}\sim 0.25$.

\subsubsection{The case $\mathcal{E}\neq 0$ and $U_{0}=0$}
\label{powerlaw}
This is the pure dilaton solution since the dilaton potential is null $U_{0}=0$. We note from (\ref{eqn:firstintbeta}) that a solution is possible only if $\mathcal{E}>0$, and now choosing $U_{0}=0$ in (\ref{eqn:betaint}) we obtain 
\begin{equation}
\beta=\sqrt{2\mathcal{E}}\; t
\quad\textnormal{and}\quad
\varphi=-\frac{1}{2}\ln {\left(\sqrt{2\mathcal{E}}\; t\right)},\quad t\geq 0.
\end{equation}
The scale factors of the external and internal dimensions are as follows:
\begin{equation}
\label{eqn:scalespowerlaw}
a=a_{0} t^{\pm\frac{1}{3}\sqrt{\frac{3n}{3+n}} \sqrt{w} }
\quad\textnormal{and}\quad
s=s_{0} t^{\mp\frac{1}{n}\sqrt{\frac{3n}{3+n}} \sqrt{w} }.
\end{equation}
This is the well-known power-law dynamics and yields the following constant deceleration parameters for the external and internal dimensions:
\begin{equation}
q_{a}= \pm 3\sqrt{\frac{3+n}{3n}}\frac{1}{\sqrt{w}}-1
\quad\textnormal{and}\quad
q_{s}= \mp n\sqrt{\frac{3+n}{3n}}\frac{1}{\sqrt{w}}-1.
\end{equation}
In this solution the expansion/contraction of the external space and contraction/expansion of the internal space start at $t=0$ and last forever. We note that the values of the deceleration parameters are determined together by the number of the internal dimensions $n$ and dilaton coupling parameter $w$ and accelerated expansion ($q_{a}<0$) is also possible for the external space. One may observe that the external space exhibits accelerated expansion, if $w>12$ for $n=1$, and if $w>3$ in the limit $n\rightarrow \infty$. On the other hand, accelerated expansion of the external space is not possible if $w\leq 3$, independent of the number of the internal dimensions. To be precise considering the present value of the deceleration parameter of the observed universe $q_{a,{\rm today}}\sim -0.81$ we find that 
\begin{equation}
w\sim 83.10 \left(1+\frac{3}{n}\right).
\end{equation}
The constraints on such power-law cosmologies $a \propto t^{\alpha}$ (equivalently $q=-1+\frac{1}{\alpha}$) from primordial nucleosynthesis are investigated in Ref. \cite{Kaplinghat99}, regardless of specific theory of gravity or the matter content of the universe by choosing $14\,{\rm Gyr}$ and $2.73\,{\rm K}$ for the age and CBR temperature of the present universe. It is found that in such power-law cosmologies, the cases $\alpha\lesssim 0.58$ ensure some primordial nucleosynthesis to take place in the early universe and the case $\alpha\approx 0.55$ can produce correct abundances of light elements. Adopting these values to our models, we find that the models for
\begin{equation}
w\lesssim 1.01 \left(1+\frac{3}{n}\right)
\end{equation}
ensure some primordial nucleosynthesis, and that the models for
\begin{equation}
\label{eqn:bbn}
w\sim 0.91 \left(1+\frac{3}{n}\right)
\end{equation}
can produce correct abundances of the light elements. Accordingly, as should be expected, the power-law dynamics cannot describe the complete picture of the universe we observe but may be valid during a particular period of the universe depending on $w$ and $n$. Moreover, the cost of representing the currently accelerated universe is very high $w$ values. Then, the question we face is whether solutions that give a more complete picture of the observed universe for a specific value of $w$ value, and hence solutions where the external space exhibits a variable deceleration parameter consistent with the observed universe can be obtained. We give promising solutions in this respect in the following sections.

\subsubsection{The case $\mathcal{E}\neq 0$ and $U_{0}\neq 0$}
\label{varyingdp}

This is the case where we obtain a time-varying deceleration parameter and hence is the one we are particularly interested in. First of all, we note from (\ref{eqn:16final}) that solutions exist only if $\mathcal{E}>0$ as we have already discussed above. Then one may observe from (\ref{eqn:betaint}) and (\ref{eqn:16final}) that we have various dynamics depending on the value of the dilaton coupling constant $w$ since it determines the behavior of the $\beta$ field. We are particularly interested in the solution where the observed universe (the 3-dimensional external space) is expanding and its deceleration parameter is a decreasing function of time, i.e. $q_{a}'<0$. One may observe from (\ref{eqn:dpbeta}) that $q_{a}'<0$ is satisfied provided that $\beta''<0$. This condition also implies from (\ref{eqn:betadiffeqn}) that $U_{0}>0$. Hence in what follows we discuss solutions for $\mathcal{E}>0$ and $U_{0}>0$ only.

We observe from (\ref{eqn:scales}) that the scale factor of the expanding external space goes either to zero or to infinity depending on the sign of the exponent as $\beta\rightarrow 0$. It follows from (\ref{eqn:firstintbeta}) that $\beta=0$ at $\beta'=\pm \sqrt{2\mathcal{E}}$. Then rewriting the integral as $\int{\frac{1}{\beta}} {\rm d}t=\int \frac{{\rm d} \ln{\beta}}{\beta'}$, we see that it goes to minus infinity (hence $a\rightarrow 0$) as $\beta'\rightarrow\sqrt{2\mathcal{E}}$ and goes to plus infinity (hence $a\rightarrow \infty$) as $\beta'\rightarrow -\sqrt{2\mathcal{E}}$ independent of the dilaton coupling constant $w$. On the other hand, the evolution trajectory of the function $\beta'$ and hence of the scale factors are determined by the dilaton coupling parameter $w$. However, note that the value of $\beta'$ at $\beta=0$ is determined only by the value of $\mathcal{E}$.  Moreover, $\beta'$ is the only function that appears in the deceleration parameters (\ref{eqn:dpbeta}). Therefore we are able to explore the initial and final values of the deceleration parameter of the external dimensions, although we cannot obtain the $\beta$ function explicitly for arbitrary $w$ values. Thus the deceleration parameter of the external dimensions when $a=0$ ($\beta'=\sqrt{2\mathcal{E}}$) and $a\rightarrow\infty$ ($\beta'=-\sqrt{2\mathcal{E}}$) from (\ref{eqn:dpbeta}) are as follows:
\begin{eqnarray}
q_{a=0}\equiv q_{a}(a=0)=3\sqrt{\frac{3+n}{3n}}\frac{1}{\sqrt{w}}-1
\quad
\textnormal{and}
\quad
q_{a= \infty}\equiv q_{a}(a=\infty)=-3\sqrt{\frac{3+n}{3n}}\frac{1}{\sqrt{w}}-1.
\end{eqnarray}
We note that the potential of the $\beta$ field is null $\mathcal{U}=0$ when $\beta'=\sqrt{2\mathcal{E}}$. Hence $q_{a=0}$ we obtained here is the same with the one we obtained for the case $\mathcal{E}> 0$ and $U_{0}=0$, i.e. the case with zero potential, in the previous section. Hence the initial ($q_{a=0}$) and the final ($q_{a=\infty}$) values of the deceleration parameter of the external space are determined by the number of the internal dimensions $n$ and the value of the dilaton parameter $w$ only. One may observe that $q_{a=0}>-1$ and $q_{a=\infty}<-1$, and hence the evolution of the external space passes through $q_{a}=-1$ (when $\beta$ reaches its maximum, i.e., $\beta'=0$) for any finite value of $w>0$.

\bigskip

\begin{figure}

\begin{minipage}[b]{0.49\linewidth}
\centering
\includegraphics[width=1\textwidth]{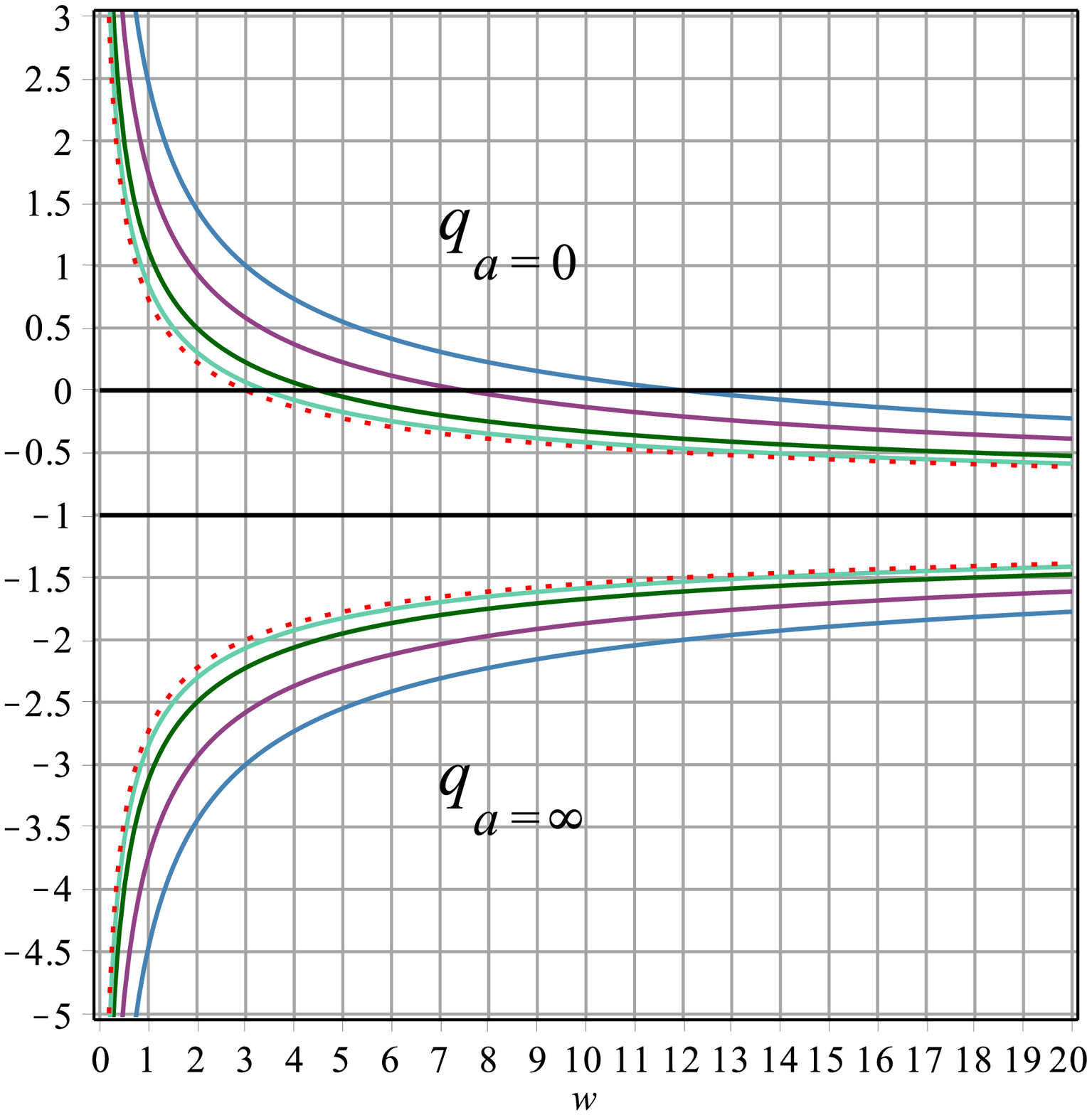}
\caption{The initial ($q_{a=0}$) and final ($q_{a=\infty}$) values of the deceleration parameter of the external space versus dilaton coupling parameter $w$ for different numbers of internal dimensions $n$. We plotted for $n=1,2,6,22$ and for $n\rightarrow \infty$ limit, in the order from the solid blue curve to the dotted red curve.}
\label{fig:constraintGw}
\end{minipage}
\hspace{0.02\linewidth}
\begin{minipage}[b]{0.49\linewidth}
\centering
\includegraphics[width=1\textwidth]{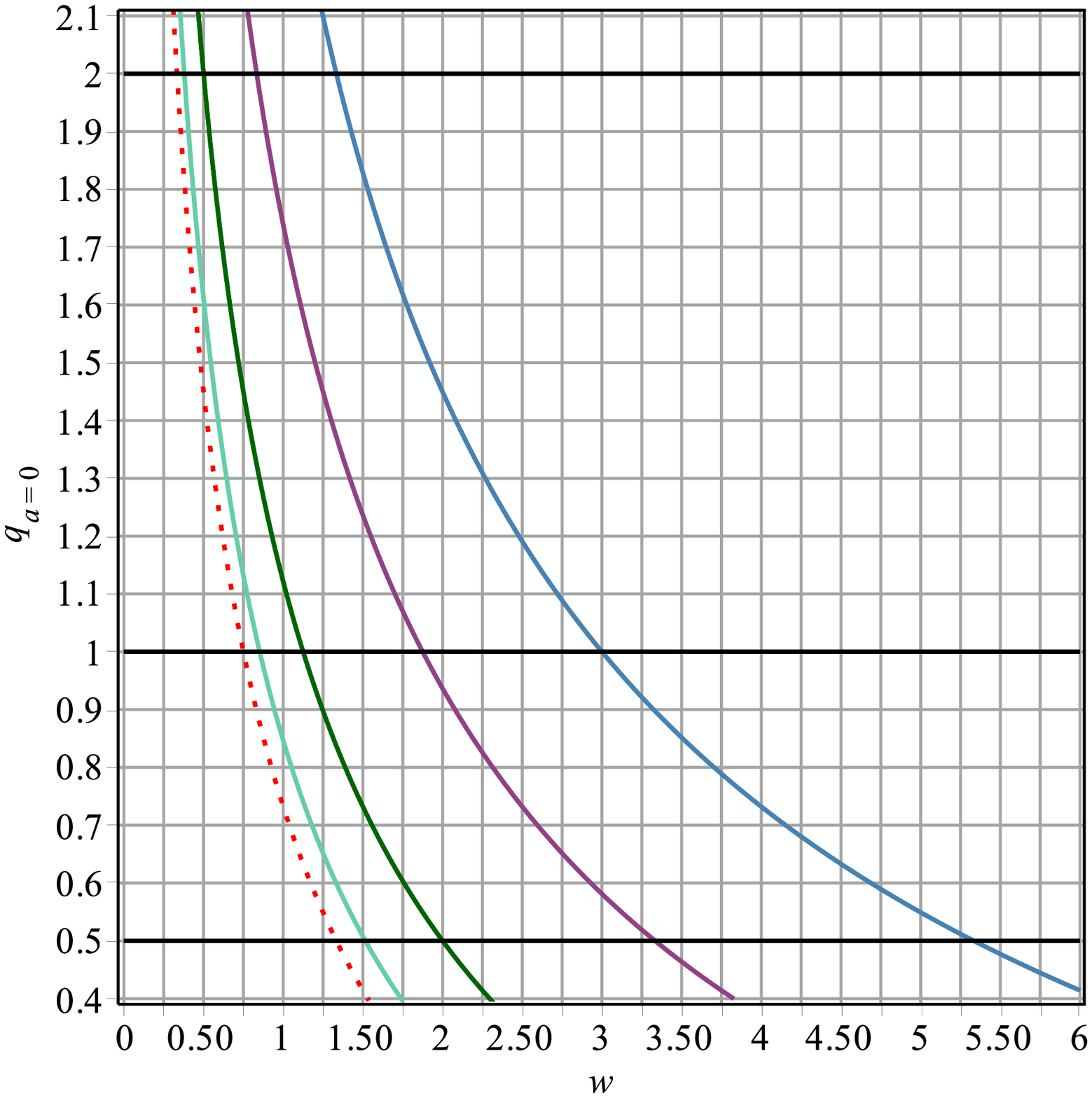}
\caption{The initial ($q_{a=0}$) value of the deceleration parameter of the external space versus dilaton coupling parameter $w$ for different numbers of internal dimensions $n$ for $0.5\leq q_{a=0}\leq 2$. $q_{a=0}=1$ corresponds to the value of the deceleration parameter in the SBBN model. We plotted for $n=1,2,6,22$ and for $n\rightarrow \infty$ limit, in the order from the solid blue curve to the dotted red curve.}
\label{fig:constraintTw}
\end{minipage}

\end{figure}

Because of the fact that we observe a universe which is evolving from a decelerated expansion phase to an accelerated expansion phase we expect at the beginning the deceleration parameter of the external space to be larger than 0 at $a=0$. Therefore, we are able to put limits on the value of $w$ with respect to the number of internal dimensions $n$:
\begin{equation}
q_{a=0}>0 \quad\Longrightarrow\quad 0<w<3+\frac{9}{n}.
\end{equation}
Thus given the number of internal dimensions, we have a certain range of allowed $w$ values that satisfy $q_{a=0}>0$ (see Fig. \ref{fig:constraintGw}). The higher the number of internal dimensions the lower the maximum value of $w$. Hence choosing the lowest value for the number of the internal dimensions $n=1$ we see that a 3-space that evolves from decelerated expansion to accelerated expansion is possible only if $w< 12$. For example, if $n=6$ then only the models with $0<w<\frac{9}{2}$ give a 3-space that starts expanding with a decelerated expansion rate. This inequality, in return, also puts constraints on the number of internal dimensions $n$. For example, given $w=6$, the number of the internal dimensions should be less than 3. On the other hand, for $w=1$ the 3-space starts expanding with a decelerated expansion rate for any number of the internal dimensions.

We can tighten the bounds on the parameters by introducing upper and lower limits on $q_{a=0}$ so as to have an idea whether a solution can accommodate the observed universe. In the conventional four dimensional cosmology in general relativity, a stiff fluid equation of state was suggested as the most promising equation of state (EoS) of matter at ultra-high densities for representing the very early universe \cite{Zeldovich62,Barrow78}. A stiff fluid can be represented with an EoS $p=\rho$, where $\rho$ and $p$ are the energy density and pressure respectively. Such an EoS gives a deceleration parameter equal to $2$ in general relativity. Motivated by this observation we choose $q_{a=0}=2$ as the upper limit for a possibly viable cosmological model. For the lower limit we can simply choose $q_{a=0}=\frac{1}{2}$, considering $\Lambda$CDM which is the simplest model that is consistent with the observations and in which universe evolves from $q_{a=0}=\frac{1}{2}$ to $q_{a=\infty}=-1$. Hence the viable values of $n$ and $w$ should satisfy the bounds
\begin{equation}
\label{eqn:tightconstraint}
\frac{1}{2} \leq q_{a=0}\leq 2 \quad \Longrightarrow \quad \frac{1}{3}+\frac{1}{n}\leq w \leq \frac{4}{3}+\frac{4}{n},
\end{equation}
which is also depicted in Fig. \ref{fig:constraintTw}. For instance, if $n=6$ viable models should satisfy $\frac{1}{2}\leq w \leq 2$. On the other hand, if $w=2$ only the models for $n\leq 6$ are viable or if $w=1$ only the models for $n\geq 2$ are viable. $w=\frac{4}{3}$ is the only value of the dilaton coupling constant that is allowed for any number of the internal dimensions, giving $q_{a=0}=2$ for $n=1$ and $q_{a=0}=\frac{1}{2}$ as $n\rightarrow\infty$.

We conclude from (\ref{eqn:tightconstraint}) (or see Fig. \ref{fig:constraintTw}) that the models for $0.33 \lesssim w \lesssim 5.33$ may describe the observed dynamics of the universe. However, besides the above constraints, for the models where $q_{a}$ is almost constant during the early the universe (so that the value of $q_{a}$ when BBN takes place will be almost the same as $q_{a=0}$), the most accurate constraints on $w$ and $n$ may be obtained through BBN. To do that we first simply consider the \textit{standard Big Bang Nucleosynthesis} (SBBN) for which it is assumed that the standard model of particle physics is valid (i.e., there are three families of neutrinos $N_{\nu}\approx 3$) and the expansion rate of the universe is governed by general relativity. The nucleosynthesis proccesses occur when the temperature ranges from $T\sim 1\, {\rm MeV}$ to $T\sim 0.1\, {\rm MeV}$ and the age of the universe varies from $t\sim 1\,{\rm s}$ to $t\sim 3\,{\rm min}$. It is further assumed that the effective equation of state of the physical content of the universe during that time interval can be described by $p=\rho/3$, which gives the expansion rate of the universe as $H_{\rm SBBN} = \frac{0.5}{t}$ (i.e., power-law expansion with a deceleration parameter value $q_{\rm SBBN}=1$) through the Friedmann equations. Therefore, we demand
\begin{equation}
\label{eqn:SBBNconstraint}
q_{a=0}= 1 \quad \Longrightarrow \quad w = \frac{3}{4}\left(1+\frac{3}{n}\right),
\end{equation}
so that our 3-space at early times mimics the expansion of the universe in SBBN model. Accordingly, the dynamics of the early universe in the SBBN can be reproduced in models for which $\frac{3}{4}\leq w \leq 3$ (see Fig. \ref{fig:constraintTw}). However, we don't have to confine our discussions to the cases $q_{a=0}= 1$. In the SBBN model, the expansion rate of the universe is given as $H_{\rm SBBN}= \frac{0.5}{t}$ under the assumption that there are three types of neutrinos $N_{\nu}\approx 3$. If on the other hand, the number of neutrino types $N_{\nu}$ is different from three, we have an altered expansion rate $H=[1+7(N_{\nu}-3)/43]^{1/2}\,H_{\rm SBBN}$ through general relativity, which in return alters the abundances of light elements (particularly of $^4$He, in contrast to D, $^3$He and $^7$Li). Because of this, it is usual to relate the observed $Y_{\rm p}$ values with the number of neutrino types \cite{Steigman05,Steigman08,Abazajian12}. The current estimates on $N_{\nu}$ from BBN in this context indicates that $N_{\nu}\sim 3-4$ is compatible with observations, with perhaps a slight preference for higher-than-standard value $N_{\nu}\approx 3$ (see \cite{Abazajian12} and references therein). On the other hand, non-standard expansion rates of the universe during the BBN may also originate due to a theory of gravity other than general relativity and the presence of extra dimensions as in our study. Although a detailed analysis of the production of light elements is out of the scope of this paper, we can make use of a good approximation for a primordial $^4$He mass fraction in the range $0.22\lesssim Y_{\rm p}\lesssim 0.27$ given by Steigman \cite{Steigman05,Steigman08} to predict $Y_{\rm p}$ values for non-standard expansion rates during BBN and then can compare our models with them by taking the observed $Y_{\rm p}$ values as our reference point. If the assumption of the SBBN model expansion rate is relaxed, both BBN and CBR will be affected and a good approximation to $Y_{\rm p}$ in this case is given as follows:
\begin{equation}
\label{eqn:Steigman}
Y_{\rm p}=0.2485\pm 0.0006+0.0016[(\eta_{10}-6)+100\,(S-1)].
\end{equation}
Here $S=H/H_{\rm SBBN}$ is the ratio of the expansion rate to the standard expansion rate and $\eta_{10}=10^{10}n_{\rm B}/n_{\gamma}$ is the ratio of baryons to photons in a comoving volume. We can safely ignore the term $\eta_{10}-6$ since observations give $\eta_{10}\sim 6$ and it is hundred times less effective than the term $S-1$. First of all we note that the SBBN model is recovered by choosing $S=1$ and the prediction for the helium mass fraction in this case is $Y_{\rm p}^{\rm SBBN}=0.2485\pm 0.0006$. On the other hand, as expected, lower the expansion rate $S<1$ the lower the helium mass fraction $Y_{\rm p}<Y_{\rm p}^{\rm SBBN}$ and higher the expansion rate $S>1$ the higher the helium mass fraction $Y_{\rm p}>Y_{\rm p}^{\rm SBBN}$. We can write this equation in a more useful form for our discussions. Using the relation between the Hubble parameter and deceleration parameter given by $q=\frac{{\rm d}}{{\rm d}t}\left(\frac{1}{H}\right)-1$ and the definition of $S$ we find $S=\frac{1+q_{\rm SBBN}}{1+q}$ where $q_{\rm SBBN}=1$ is the value of the deceleration parameter assume in the SBBN model. Accordingly, the equation given in (\ref{eqn:Steigman}) can be written in terms of $q_{a=0}$ or in terms of $w$ and $n$ as follows:
\begin{equation}
\label{eqn:qSteigman}
Y_{\rm p}=0.2485\pm 0.0006+0.16\,\frac{1-q_{a=0}}{1+q_{a=0}}=0.2485\pm 0.0006+0.16\left(-1+\frac{2}{3}\sqrt{\frac{3wn}{3+n}}\right).
\end{equation}
We should note that this relation will be valid for our models as long as $q_{a=0}$ is almost the same with the $q_{a}$ at time of the BBN. Now, using (\ref{eqn:qSteigman}), we can calculate $Y_{\rm p}$ values according to $w$ and $n$ and then compare them with the $Y_{\rm p}^{\rm SBBN}=0.2485\pm 0.0006$ and the observed $Y_{\rm p}$ values.
\begin{figure}[h]
\centering
\includegraphics[width=0.5\textwidth]{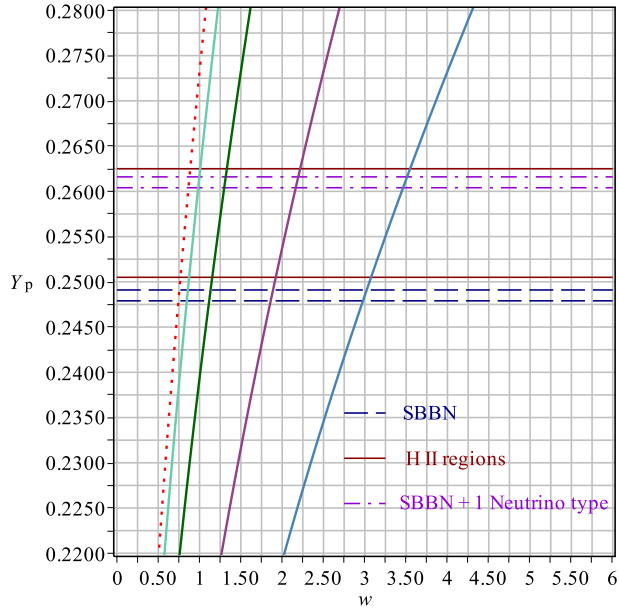}
\caption{The predicted $^4$He mass fraction ($Y_{\rm p}$) versus $w$ for $n=1,2,6,22$ and as $n\rightarrow\infty$, in the order from the solid blue cruve to the dotted red curve. The SBBN prediction $Y_{\rm p}^{\rm SBBN}=0.2485\pm 0.0006$ \cite{Steigman08} corresponds to the region between the two dashed horizontal lines. Observationally inferred recent value $Y_{\rm p}=0.2565\pm 0.0060$ \cite{Izotov10}, from the hydrogen and helium recombination lines from low-metallicity extragalactic H II regions that is most useful observations for BBN, corresponds to the region between the two solid horizontal lines. The value $Y_{\rm p}^{\rm SBBN}=0.2610\pm 0.0006$ \cite{Steigman08} is the predicted $^4$He mass fraction when an extra neutrino type is included to the SBBN model and corresponds to the region between two dashed-dotted horizontal lines.}
\label{fig:helium}
\end{figure}
The most useful observations for BBN are those of the hydrogen and helium recombination lines from low-metallicity, extragalactic H II regions. Two recent results from such observations are $Y_{\rm p}=0.2565\pm 0.0060$ \cite{Izotov10} and $Y_{\rm p}=0.2561\pm 0.0108$ \cite{Aver10}. We plot $Y_{\rm p}$ versus $w$ for $n=1,2,6,22$ and as $n\rightarrow\infty$ and compare them with the SBBN prediction $Y_{\rm p}^{\rm SBBN}$ and observationally inferred $Y_{\rm p}$ value in Fig. \ref{fig:helium}. We pay particular attention to the predicted $Y_{\rm p}$ values for two interesting cases and would like to check them immediately. For the case $w=1$ and $n=6$ that correponds to anomaly-free superstring theory, the predicted helium mass fraction is $Y_{\rm p}=0.2393\pm 0.0006$, which is less than both the SBBN and observed $Y_{\rm p}$ values. On the other hand, for the case $w=1$ and $n=22$ that correponds to anomaly-free bosonic string theory, the predicted helium mass fraction is $Y_{\rm p}=0.2618\pm 0.0006$, which agrees with the observed $Y_{\rm p}$ values (for instance with $Y_{\rm p}=0.2561\pm 0.0108$) even better than the SBBN prediction $Y_{\rm p}^{\rm SBBN}=0.2485\pm 0.0006$. This is a very promising result. However, of course, we should further examine two other points. First we examine if the deceleration parameter of the 3-space varies very slowly in the early universe to see whether our approximation is valid or not. Second we examine if the age of the universe was less than the lifetime of the free neutrons ($\tau_{\rm n}\sim887\,{\rm s}$) when the CBR temperature reaches  $T\sim 80\,{\rm keV}$. Otherwise the BBN model would not work  properly and then our above predictions would not be valid. Both of these points can be easily checked, if the scale factors are obtained explicitly which shall be achieved for the cases for $w=1$ and $w=\frac{1}{2}$ below. In short we have various models according to the value of the dilaton coupling constant $w$ and the number of the internal dimensions $n$ and can even have cosmological models in which the effective four dimensional universe can sustain favorable conditions for a successful primordial nucleosynthesis in the early universe and evolves from a decelerated expansion phase to an accelerated expansion phase consistently with the data coming from cosmic microwave background and supernova observations.

We would also like to comment on the future of the universe in our models. It is well-known that the deceleration parameter values lower than $-1$, i.e. expansion rates faster than the exponential expansion, cause a scale factor to expand to infinitely large values in finite time, a behavior known as the Big Rip \cite{Caldwell03}. Hence in the models given in this section the external space will expand to infinitely large values in finite time since $q_{a=\infty}<-1$ for any combination of $n$ and $w$. The same result can also be derived from (\ref{eqn:Hbeta}). $H_{a}$ starts with infinitely large values when $a=0$ ($\beta=0$ and $\beta'=\sqrt{2\mathcal{E}}$), decreases and reaches its minimum when $\beta=\beta_{\rm max}$ and starts to increase and becomes infinitely large as $a\rightarrow\infty$ ($\beta=0$ and $\beta'=-\sqrt{2\mathcal{E}}$).

\bigskip
Up to this point we have discussed the cases for $\mathcal{E}> 0$ and $U_{0}>0$ qualitatively because the function $\beta$ is, in general, cannot be expressed in terms of known functions. However, it is expressible in terms of known functions and $\int {\frac{1}{\beta}} {\rm d} t$ is integrable so as to obtain the scale factors explicitly in two cases:
\begin{itemize}

\item
Case $w=1$ ($\lambda=0$):
We let  $w=1$ in (\ref{eqn:betadiffeqn}) so that
\begin{equation}
\beta''=-U_{0}\beta,
\end{equation}
which is the familiar simple harmonic oscillator equation and the solution of (\ref{eqn:betaint}) gives
\begin{equation}
\label{eqn:betap1}
\beta=\sqrt{\frac{2\mathcal{E}}{U_{0}}}\sin (\sqrt{U_{0}}\;t),\quad 0\leq t \leq \frac{\pi}{\sqrt{U_{0}}}.
\end{equation}

\item
Case $w=\frac{1}{2}$ ($\lambda=2$):
We let $w=\frac{1}{2}$ in (\ref{eqn:betadiffeqn}) so that
\begin{equation}
\beta''=-U_{0},
\end{equation}
which is similar to the equation of motion of a particle under a constant attractive force in classical physics, and the solution of (\ref{eqn:betaint}) gives
\begin{equation}
\label{eqn:betap1b2}
\beta=\frac{\mathcal{E}}{U_{0}}-\frac{1}{2}U_{0}\;t^{2}, \quad -\frac{\sqrt{2\mathcal{E}}}{U_{0}}\leq t \leq \frac{\sqrt{2\mathcal{E}}}{U_{0}}.
\end{equation}
\end{itemize}
In what follows, we discuss each of these two exact solutions in further detail.

%%%%%%%%%%%%%%%%%%%%%%%%%%%%%%%%%%%%%

\section{The model for $w=1$ ($\lambda=0$)}
\label{sec:w1}

The special choice $w=1$ reduces the action (\ref{eqn:action}) to the gravi-dilaton string effective action in $(1+3+n)$-dimensions and an explicit solution will be provided for $\mathcal{E}>0$ and $U_{0}>0$. Now, using (\ref{eqn:betap1}) in (\ref{eqn:beta}) we obtain the dilaton field as follows:
\begin{equation}
\varphi=-\frac{1}{2}\ln{\left(\sqrt{\frac{2\mathcal{E}}{U_{0}}}\sin (\sqrt{U_{0}}\;t)\right)},\quad 0\leq t \leq \frac{\pi}{\sqrt{U_{0}}}.
\end{equation}
Because $\lambda=0$ for $w=1$, the dilaton field yields a constant self-interaction potential
\begin{equation}
U=U_{0}.
\end{equation}
The scalar curvature of the higher dimensional spacetime
\begin{equation}
\mathcal{R} =\frac{U_{0}}{\sin^{2}\left(\sqrt{U_{0}}\; t\right)}.
\end{equation}
We obtain the scale factors of the external and internal spaces as
\begin{equation}
\label{eqn:scalesw1}
a=a_{1}\,e^{\frac{1}{3}\sqrt{\frac{3n}{3+n}}\,{\rm{arctanh}}\left(-\cos(\sqrt{U_{0}}\;t)\right)}
\quad\textnormal{and}\quad
s=s_{1}\,e^{-\frac{1}{n}\sqrt{\frac{3n}{3+n}}\,{\rm{arctanh}}\left(-\cos(\sqrt{U_{0}}\;t)\right)},
\end{equation}
respectively, where $a_{1}$ and $s_{1}$ are integration constants that obey ${a_{1}}^3{s_{1}}^n=V_{0}$. The Hubble parameters of the external and internal spaces are
\begin{equation}
H_{a}=\frac{1}{3}\sqrt{\frac{3n}{3+n}}\;\frac{\sqrt{U_{0}}}{\sin(\sqrt{U_{0}}\; t)}
\quad\textnormal{and}\quad
H_{s}=-\frac{1}{n}\sqrt{\frac{3n}{3+n}}\;\frac{\sqrt{U_{0}}}{\sin(\sqrt{U_{0}}\; t)}.
\end{equation}
The deceleration parameters of the external and internal spaces are
\begin{equation}
q_{a}=-1+3\sqrt{\frac{3+n}{3n}}\;\cos(\sqrt{U_{0}}\; t)
\quad\textnormal{and}\quad
q_{s}=-1-n\sqrt{\frac{3+n}{3n}}\;\cos(\sqrt{U_{0}}\; t).
\end{equation}
We also give the jerk parameter for the external space only, since it is important while discussing the model from the observational point of view:
\begin{equation}
j_{a}=\left(3+\frac{9}{n}\right) \cos^{2}(\sqrt{U_{0}}\; t)+9\sqrt{\frac{3+n}{3n}}\cos(\sqrt{U_{0}}\; t)+\frac{9}{n}+4.
\end{equation}

\begin{figure}

\begin{minipage}[b]{0.49\linewidth}
\centering
\includegraphics[width=1\textwidth]{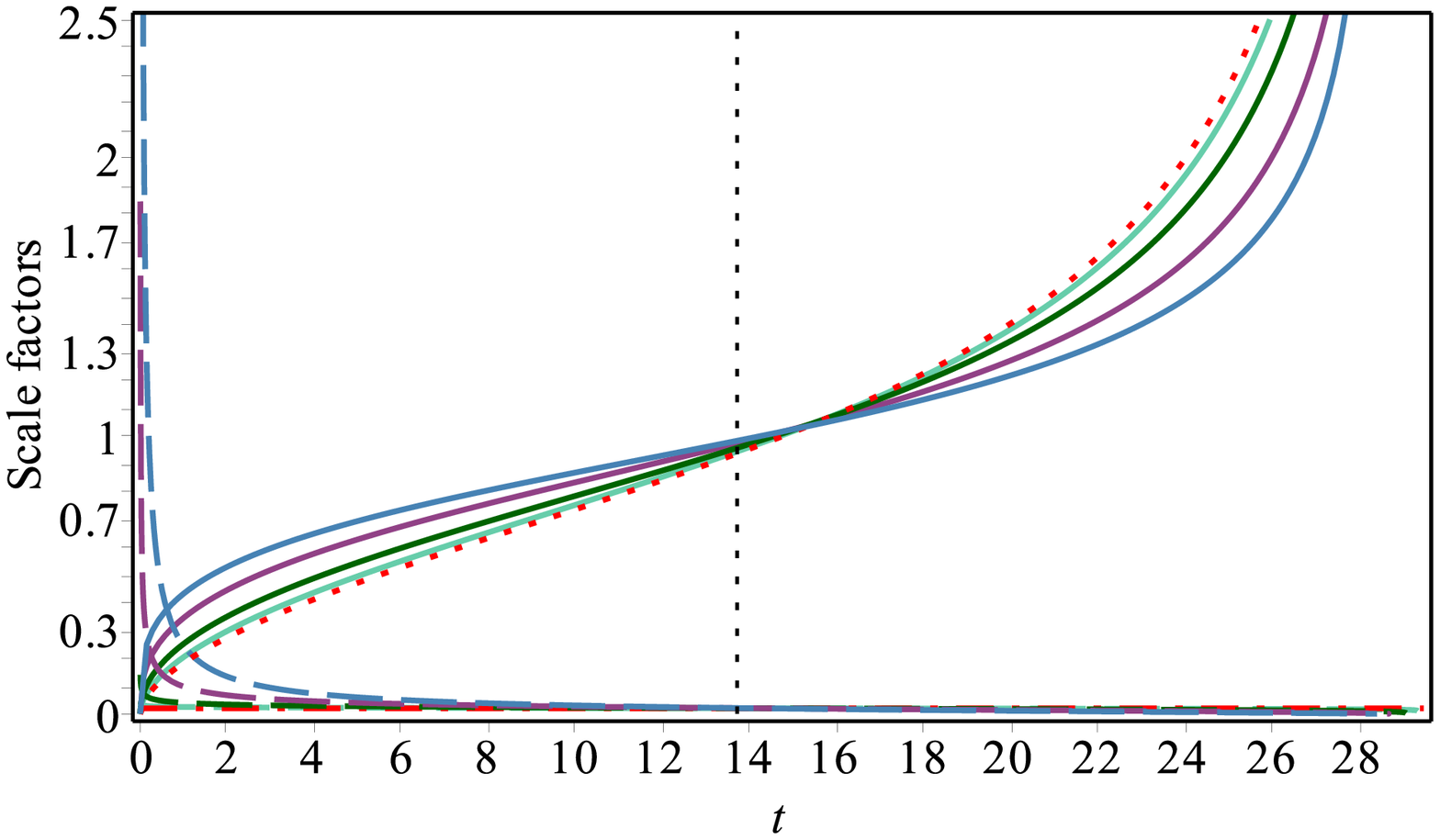}
\caption{Scale factors of the external and internal spaces versus cosmic time $t$ for different numbers of internal dimensions. We plotted for $n=1,2,6,22$ and for $n\rightarrow \infty$ limit, in the order from the solid blue curve to the dotted red curve.}
\label{fig:sfw1}
\end{minipage}
\hspace{0.01\linewidth}
\begin{minipage}[b]{0.49\linewidth}
\centering
\includegraphics[width=1\textwidth]{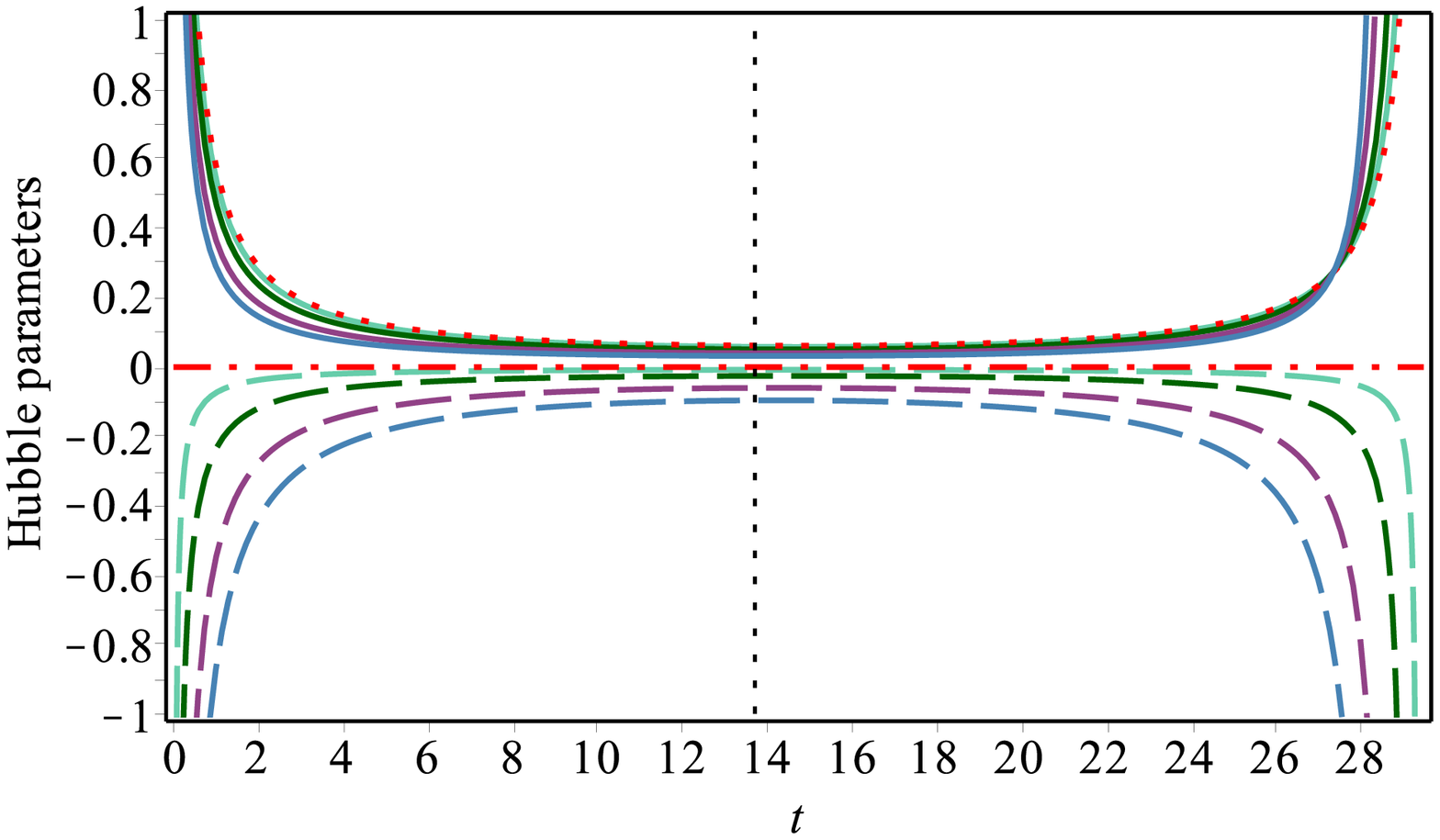}
\caption{Hubble parameters of the external and internal spaces versus cosmic time $t$ for different numbers of internal dimensions. We plotted for $n=1,2,6,22$ and for $n\rightarrow \infty$ limit, in the order from the solid blue curve to the dotted red curve.}
\label{fig:hw1}
\end{minipage}

\begin{minipage}[b]{0.49\linewidth}
\centering
\includegraphics[width=1\textwidth]{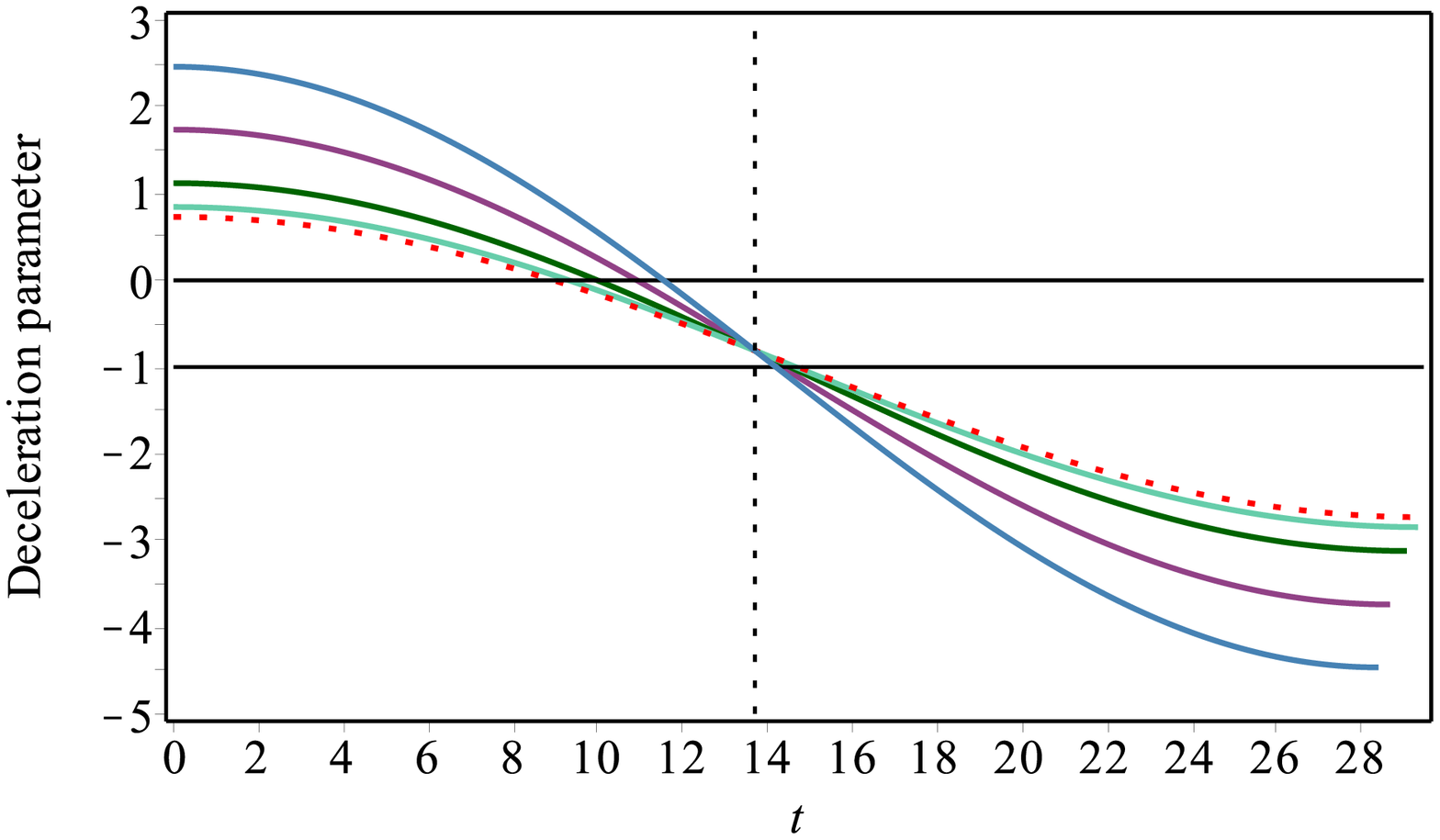}
\caption{Deceleration parameters of the external space versus cosmic time $t$ for different numbers of internal dimensions. We plotted for $n=1,2,6,22$ and for $n\rightarrow \infty$ limit, in the order from the solid blue curve to the dotted red curve.}
\label{fig:qw1}
\end{minipage}
\hspace{0.01\linewidth}
\begin{minipage}[b]{0.49\linewidth}
\centering
\includegraphics[width=1\textwidth]{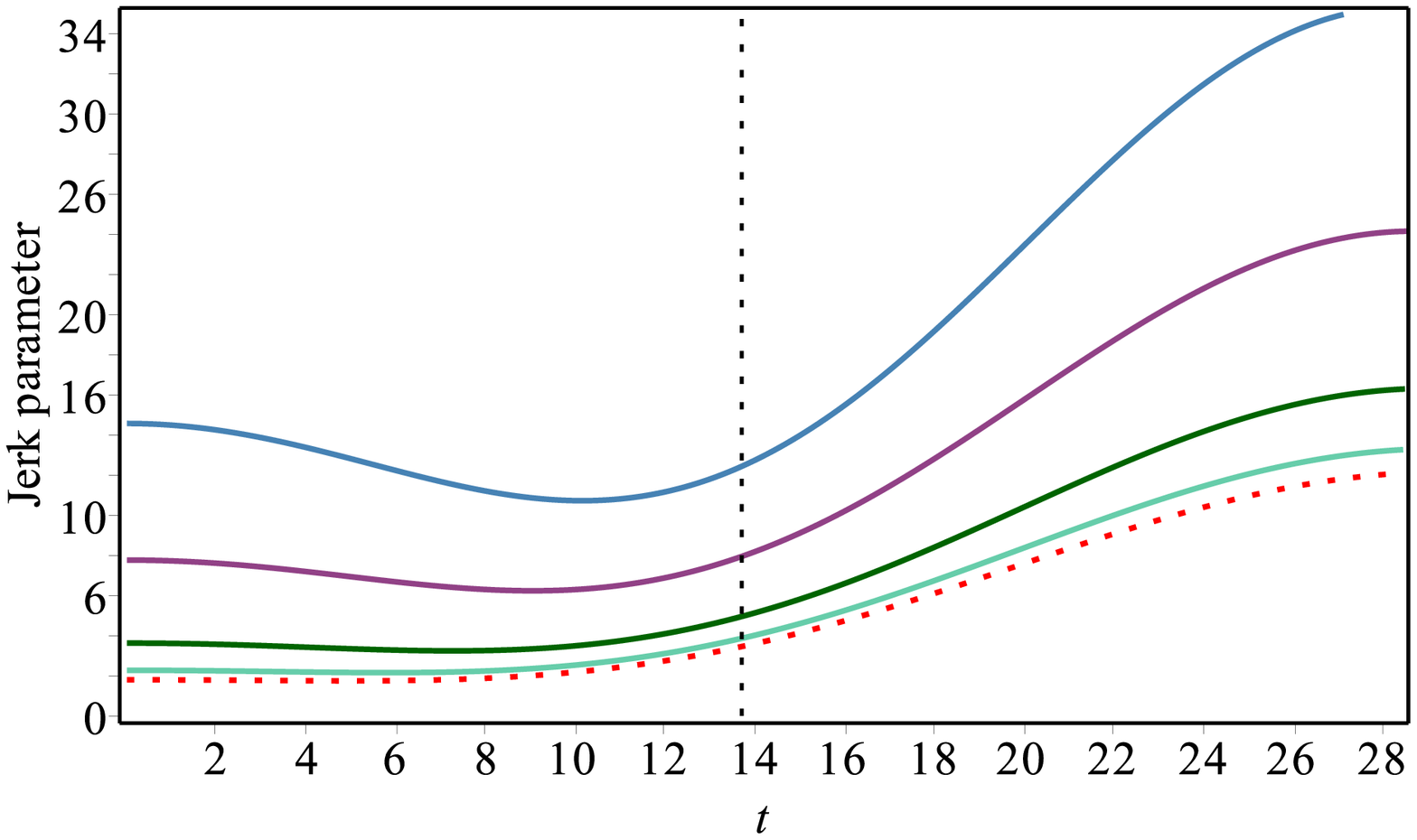}
\caption{Jerk parameters of the external space versus cosmic time $t$ for different numbers of internal dimensions. We plotted for $n=1,2,6,22$ and for $n\rightarrow \infty$ limit, in the order from the solid blue curve to the dotted red curve.}
\label{fig:jw1}
\end{minipage}
\end{figure}

\begin{figure}
\centering
\includegraphics[width=0.5\textwidth]{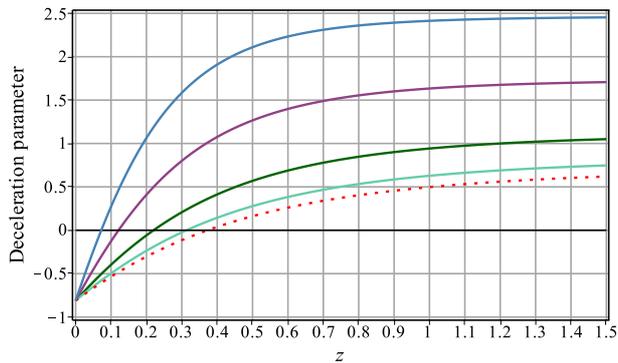}
\caption{Deceleration parameters of the external space versus cosmic redshift $z$ for different numbers of internal dimensions. We plotted for $n=1,2,6,22$ and for $n\rightarrow \infty$ limit, in the order from the solid blue curve to the dotted red curve.}
\label{fig:qw1z}

\end{figure}

The evolution of the universe starts with a zero size external space and infinitely large internal space at $t=0$.  The external space that we observe today expands until it becomes infinitely large at a finite time $t_{\rm end}=\frac{\pi}{\sqrt{U_{0}}}$, while the internal space contracts to unobservable scales and then reaches zero size at $t_{\rm end}$. The external space enters into the accelerated expansion phase at $t_{\rm acc}=\frac{1}{\sqrt{U_{0}}}\arccos\left(\frac{1}{3}\sqrt{\frac{3n}{3+n}} \right)$, passes over into the super-exponential expansion phase at $t_{\rm se}=\frac{\pi}{2\sqrt{U_{0}}}$ and then reaches infinitely large expansion rate at $t=t_{\rm end}$. We note that all the cosmological parameters are exactly determined by the values of $U_{0}$ and $n$ only; in particular $U_{0}$ determines the lifetime of the universe, $n$ determines the initial and final values of the deceleration parameters and they together determine the time of the onset of the acceleration of the external space. Hence, in principle, we are able to obtain the number of internal dimensions $n$ and $U_{0}$ upon obtaining the values of the deceleration parameter at two different times or cosmic redshifts from observations. However, in this paper, we won't undertake an observational study but discuss the dynamics for different numbers of internal dimensions simply by considering the present age of the universe and the present value of the deceleration parameter that are consistent with various observational studies, in particular concerning model independent studies. Hence, with the choice $q_{a,0}=-0.81$ \cite{Visser04,Rapetti07,Cattoen08,Wang09,Vitagliano10,Capozziello11,Xia12} and $t_{\rm today}=13.7\,{\rm Gyr}$ \cite{Komatsu11} considering the latest observations, we are able to depict the dynamics of the universe for different numbers of internal dimensions $n$. Accordingly, we plot the scale factors of the external and internal dimensions in Fig. \ref{fig:sfw1} and the Hubble parameters in Fig. \ref{fig:hw1} for $n=1,\;2,\;6,\;22$ and also for $n\rightarrow \infty$ limit. We plot the dimensionless deceleration parameter of the external dimensions in Fig. \ref{fig:qw1} and the jerk parameter in Fig. \ref{fig:jw1}. The evolution of the deceleration parameter with the cosmic redshift is important to have an idea about the evolution of the observed universe. Hence, we also give the deceleration parameter of the external space versus the cosmic redshift $z=-1+\frac{a_{z=0}}{a}$ (where $a_{z=0}$ is the present value of
the scale factor of the external space)
\begin{equation}
q_{a}(z)=-1+3\sqrt{\frac{3+n}{3n}}\tanh\left(3\sqrt{\frac{3+n}{3n}}\ln(1+z)+{\rm{arctanh}}\left(\frac{1}{3}\sqrt{\frac{3n}{3+n}}(q_{0}+1)\right) \right)
\end{equation}
and plot the deceleration parameter of the external space in terms of cosmic redshift $z$ in Fig. \ref{fig:qw1z}. We summarize, in Table \ref{table:1}, the plots by giving the cosmic redshift ($z_{\rm acc}$) and the time that passed since the time of the onset of acceleration ($13.7\,{\rm Gyr}-t_{\rm acc}$), the time of the Big Rip ($t_{\rm end}$) and the initial ($q_{a,{\rm int}}$, $j_{a,{\rm int}}$), present ($q_{a,0}$, $j_{a,0}$) and final ($q_{a,{\rm end}}$, $j_{a,{\rm end}}$) values of the deceleration and jerk parameters.

We observe that while the number of internal dimensions alters the expansion rate of the external dimensions slightly, it alters the contraction rate of the internal dimensions drastically. From Table \ref{table:1}, one may observe that it is more likely that our model accommodates the observed dynamics of the universe when the number of internal dimensions is higher. As we discussed on Fig. \ref{fig:constraintTw} we do not expect a successful model for $w=1$ and $n=1$ and hence these values can be excluded, since $q_{a,{\rm int}}=2.46>2$. Indeed, one may check from Table \ref{table:1} that in this case the transition to the accelerated expansion is too recent $z_{\rm acc}=0.07$ and the value of the jerk parameter is too high $j_{a,0}=12.47$. In the case for $n=2$ also, the transition to the accelerated expansion is too recent, although the values $q_{a,{\rm int}}=1.74<2$ and $j_{a,0}=7.97$ may be acceptable. We would like to discuss the cases $n=6$ ($1+3+n=10$) and $n=22$ ($1+3+n=26$) together. With the choice $w=1$ our action reduces to the gravi-dilaton action of the anomaly-free superstring theory for $n=6$ and of the anomaly free bosonic string theory for $n=22$. The deceleration parameter of the external space that reaches $q_{a,0}=-0.81$ at the present age of the universe starts with a value $q_{a,{\rm int}}=1.12$ for $n=6$ and $q_{a,{\rm int}}=0.85$ for $n=22$. The jerk parameter, on the other hand, evolves from $j_{a,{\rm int}}=3.64$ at $t=0$ to $j_{a,0}=4.97$ at $t_{\rm today}=13.7\,{\rm Gyr}$ for $n=6$ and from $j_{a,{\rm int}}=2.28$ at $t=0$ to $j_{a,0}=3.88$ at $t_{\rm today}=13.7\,{\rm Gyr}$ for $n=22$. These values can accommodate the observed dynamics of the universe \cite{Visser04,Rapetti07,Cattoen08,Wang09,Vitagliano10,Capozziello11,Xia12}. On the other hand, the transition to the accelerated expansion occurs more recently $z_{\rm acc}=0.22$ in the case for $n=6$ to be compared with the value $z_{\rm acc}=0.31$ in the case for $n=22$, with the latter case accommodating the observations better \cite{Alam04,Sahni06,Tasos09,Lima10,Cunha09,Li11}.

Regarding the early times of the universe in our models, we should first assure the universe to be effectively four dimensional during the BBN. Accordingly, we give the maximum size of the internal dimensions today $s_{0}$ provided $s_{\rm NS}\lesssim l_{\rm proton}$ in Table \ref{table:1}. One may check that the the deceleration parameter of the 3-space at the time scales of BBN can be taken as $q_{a=0}$ in this solution and hence the equation (\ref{eqn:qSteigman}) we give for predicting the helium mass fraction can safely be used. We find that the helium mass fraction predicted by SBBN $Y_{\rm p}^{\rm SBBN}=0.2485\pm 0.0006$ can be reproduced if $n=9$. The predicted helium mass fraction in the range $n=10-24$ are consistent with the inferred helium mass fraction $Y_{\rm p}=0.2565\pm 0.0060$ from observations. We give the predicted helium mass fraction $Y_{\rm p}$ values for $n=1,\,2,\,6,\,22$ and as $n\rightarrow\infty$ values using (\ref{eqn:qSteigman}) in Table \ref{table:1}. We note that among the cases given in Table \ref{table:1} the helium mass fraction is consistent with the observational value $Y_{\rm p}=0.2565\pm 0.0060$ only for the case $n=22$ with a value $Y_{\rm P}=0.2618\pm 0.0006$. We also calculated the time $t_{T\sim 80\,{\rm keV }}$ and give in Table \ref{table:1}, to see whether the condition $t_{T\sim 80\,{\rm keV }}\lesssim 887\,{\rm s}$ is satisfied to ensure BBN to work properly and confirm that this condition is satisfied for all $n$ values. It is interesting that in the limit as $n\rightarrow\infty$, the model approaches right to the  $t\lesssim887\,{\rm s}$ bound with a value $884\,{\rm s}$.
\begin{table}\footnotesize
  \caption{Cosmological parameters for the case $w=1$ ($\lambda=0$) for different numbers of internal dimensions $n$ assuming $q_{a,0}=-0.81$ at $t_{\rm today}=13.7\,{\rm Gyr}$.}
  \label{table:1}
\begin{center}
\begin{tabular}{cc|c|c|c|c|c|c|c|c|l}
\cline{2-6}
\multicolumn{1}{l|}{} & \multicolumn{1}{c|}{$n=1$} & \multicolumn{1}{c|}{$n=2$}  & \multicolumn{1}{c|}{$n=6$} & \multicolumn{1}{c|}{$n=22$} & \multicolumn{1}{c|}{$n\rightarrow \infty$} \\[3pt]
\cline{2-6}
\cline{1-6}
\multicolumn{1}{|l|}{$z_{\rm acc}$ } & 0.07  & 0.12  & 0.22 & 0.31 & 0.37 \\[3pt]
\multicolumn{1}{|l|}{$13.7\,{\rm Gyr}-t_{\rm acc}$} & 2.15 Gyr & 2.78 Gyr & 3.71 Gyr & 4.38 Gyr  & 4.74 Gyr\\[3pt]
\multicolumn{1}{|l|}{$t_{\rm end}$} & 28.39 Gyr & 28.67 Gyr & 29.06 Gyr & 29.32 Gyr & 29.46 Gyr \\[3pt]
\cline{1-6}
\multicolumn{1}{|l|}{$q_{a,{\rm int}}$, $j_{a,{\rm int}}$} & 2.46, 14.61  & 1.74, 7.78 & 1.12, 3.64 & 0.85, 2.28 & 0.73, 1.80 \\[3pt]
\multicolumn{1}{|l|}{$q_{a,0}$, $j_{a,0}$} & -0.81, 12.47 & -0.81, 7.97 & -0.81, 4.97   & -0.81, 3.88 & -0.81, 3.47 \\[3pt]
\multicolumn{1}{|l|}{$q_{a,{\rm end}}$, $j_{a,{\rm end}}$} & -4.46, 35.39 & -3.74, 24.22 & -3.12, 16.36 & -2.85, 13.36 & -2.73, 12.20 \\[3pt]
\cline{1-6}
\multicolumn{1}{|l|}{$s_{0}$} & $<10^{-43}\,{\rm m}$ & $<10^{-30}\,{\rm m}$ & $<10^{-21}\,{\rm m}$ & $<10^{-18}\,{\rm m}$ & $<10^{-16}\,{\rm m}$ \\[3pt]
\multicolumn{1}{|l|}{$Y_{\rm p}$ ($^4$He mass fraction)} & $0.1809\pm 0.0006$   & $0.2053\pm 0.0006$ & $0.2393\pm 0.0006$ & $0.2618\pm 0.0006$ & $0.2732\pm 0.0006$ \\[3pt]
\multicolumn{1}{|l|}{$t_{T\sim 80\,{\rm keV}}$} & $10^{-12}\, {\rm s}<887\,{\rm s}$   & $10^{-6}\, {\rm s}<887\,{\rm s}$ & $10^{-1}\, {\rm s}<887\,{\rm s}$ & $94\, {\rm s}<887\,{\rm s}$ & $884\, {\rm s} <887\,{\rm s}$ \\[3pt]

\cline{1-6}
\end{tabular}
\end{center}
\end{table}

We would like to end this section by a comment on an interesting coincidence. There is an ongoing discussion on the possibility of the presence of a fourth neutrino type (sterile neutrino). One may consult Ref. \cite{Abazajian12} for a recent and exhaustive review and for a list for the inferred number of the neutrino types from astrophysical and cosmological observations. As we discussed above, inclusion of an extra neutrino type into the SBBN model (in which it is assumed that there are three neutrino types) would increase the expansion rate of the universe at the time of the neutron-proton freeze-out, consequently giving rise to a higher  helium mass fraction than the one predicted by the SBBN model. Non-standard early universe expansion rate due to the number of neutrino types is parametrized with $S=(1+7\Delta N_{\nu}/43)^{\frac{1}{2}}$ where $\Delta {N_{\nu}}=N_{\nu}-3$ being the deviation from the number of the neutrino types in the standard particle physics \cite{Steigman05}. Using this equation in (\ref{eqn:Steigman}) by considering one extra neutrino type $\Delta{N_{\nu}}=1$, i.e. $N_{\nu}=4$, we find $Y_{\rm p}=0.2610\pm 0.0006$. It is quite interesting that this value is almost the same as $Y_{\rm P}=0.2618\pm 0.0006$ that we predicted when $n=22$ in our model. It is even more interesting when we recall that $w=1$ and $n=22$ is a very particular combination; the combination that reduces our action to the gravi-dilaton action of the anomaly-free bosonic string theory. 

%%%%%%%%%%%%%%%%%%%%%%%%%%%%%%%%%%%%%%%%%%%%%%

\section{The model for $w=\frac{1}{2}$ ($\lambda=2$)}
\label{sec:w1b2}

In this section we give an exact solution for the case $w=\frac{1}{2}$ ($\lambda=2$) that is valid provided that $\mathcal{E}>0$ and $U_{0}>0$. For convenience we redefine the time parameter as $t\rightarrow t-\frac{\sqrt{2\mathcal{E}}}{U_{0}}$ so as to set $a=0$ at $t=0$. Now, using (\ref{eqn:betap1b2}) in (\ref{eqn:beta}) we obtain the dilaton field as follows:
\begin{equation}
\label{eqn:27}
\varphi=-\frac{1}{2} \ln{\left(-\frac{U_{0}}{2}\;t^2+\sqrt{2\mathcal{E}}\;t\right)}.
\end{equation}
The self-interaction potential of the dilaton field
\begin{equation}
U=\frac{U_{0}}{-\frac{U_{0}}{2}\; t^2+\sqrt{2\mathcal{E}}\;t}.
\end{equation}
The scalar curvature of the higher dimensional spacetime
\begin{equation}
\mathcal{R} =\frac{1}{2} \frac{{2\mathcal{E}}}{\left(\frac{U_{0}}{2}\; t^2-\sqrt{2\mathcal{E}}\;t\right)^2}.
\end{equation}
We obtain the scale factors of the external and internal dimensions as follows:
\begin{equation}
\label{eqn:scalesw1b2}
a=a_{2}\,e^{\frac{\sqrt{2}}{3}\frac{\sqrt{3n}}{\sqrt{3+n}}\,{\rm{arctanh}}\left(\frac{U_{0}}{\sqrt{2\mathcal{E}}}\;t-1\right)}
\quad\textnormal{and}\quad
s=s_{2}\,e^{-\frac{\sqrt{2}}{n}\frac{\sqrt{3n}}{\sqrt{3+n}}\,{\rm{arctanh}}\left(\frac{U_{0}}{\sqrt{2\mathcal{E}}}\;t-1\right)},
\end{equation}
respectively, where $a_{2}$ and $s_{2}$ are integration constants that obey ${a_{2}}^3{s_{2}}^2=V_{0}$. The Hubble parameters of the external and internal dimensions are
\begin{equation}
H_{a}=\frac{2}{3}\sqrt{\frac{3n}{3+n}}\;\frac{\sqrt{\mathcal{E}}}{-U_{0}\;t^2+2\sqrt{2\mathcal{E}}\;t}
\quad\textnormal{and}\quad
H_{s}=-\frac{2}{n}\sqrt{\frac{3n}{3+n}}\;\frac{\sqrt{\mathcal{E}}}{-U_{0}\;t^2+2\sqrt{2\mathcal{E}}\;t}.
\end{equation}
The deceleration parameters of the external and internal dimensions are
\begin{equation}
q_{a}=-3\sqrt{\frac{3+n}{3n}}\frac{U_{0}}{\sqrt{\mathcal{E}}} \;t+3\sqrt{2}\sqrt{\frac{3+n}{3n}}-1
\quad\textnormal{and}\quad
q_{s}=n\sqrt{\frac{3+n}{3n}}\frac{U_{0}}{\sqrt{\mathcal{E}}}\; t-n\sqrt{2}\sqrt{\frac{3+n}{3n}}-1.
\end{equation}
We also give the jerk parameter for the external dimensions only, since it is important while discussing the model from the observational point of view:
\begin{equation}
j=\frac{\dddot{a}}{aH_{a}^3}=\frac{9}{2}\frac{U_{0}^2}{\mathcal{E}}\frac{3+n}{n}\;t^2
+\left(3\sqrt{6}\sqrt{\frac{3+n}{n}}-\frac{54}{n}-18\right)\frac{U_{0}}{\sqrt{2\mathcal{E}}}\;t
-3\sqrt{6}\sqrt{\frac{3+n}{n}}+\frac{36}{n}+13.
\end{equation}

\begin{figure}

\begin{minipage}[b]{0.49\linewidth}
\centering
\includegraphics[width=1\textwidth]{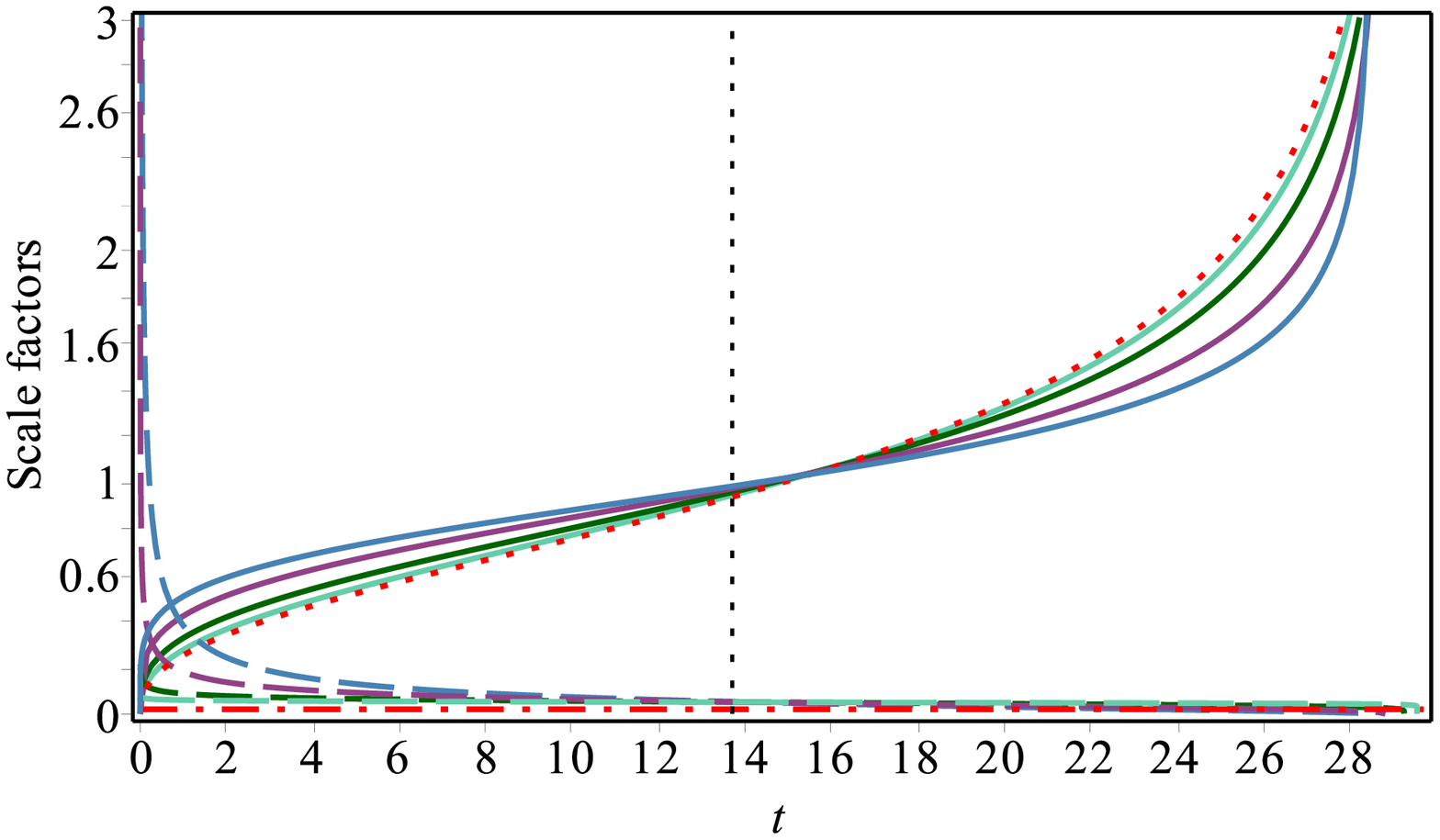}
\caption{Scale factors of the external and internal spaces versus cosmic time $t$ for different numbers of internal dimensions. We plotted for $n=1,2,6,22$ and for $n\rightarrow \infty$ limit, in the order from the solid blue curve to the dotted red curve.}
\label{fig:sfw1b2}
\end{minipage}
\hspace{0.01\linewidth}
\begin{minipage}[b]{0.49\linewidth}
\centering
\includegraphics[width=1\textwidth]{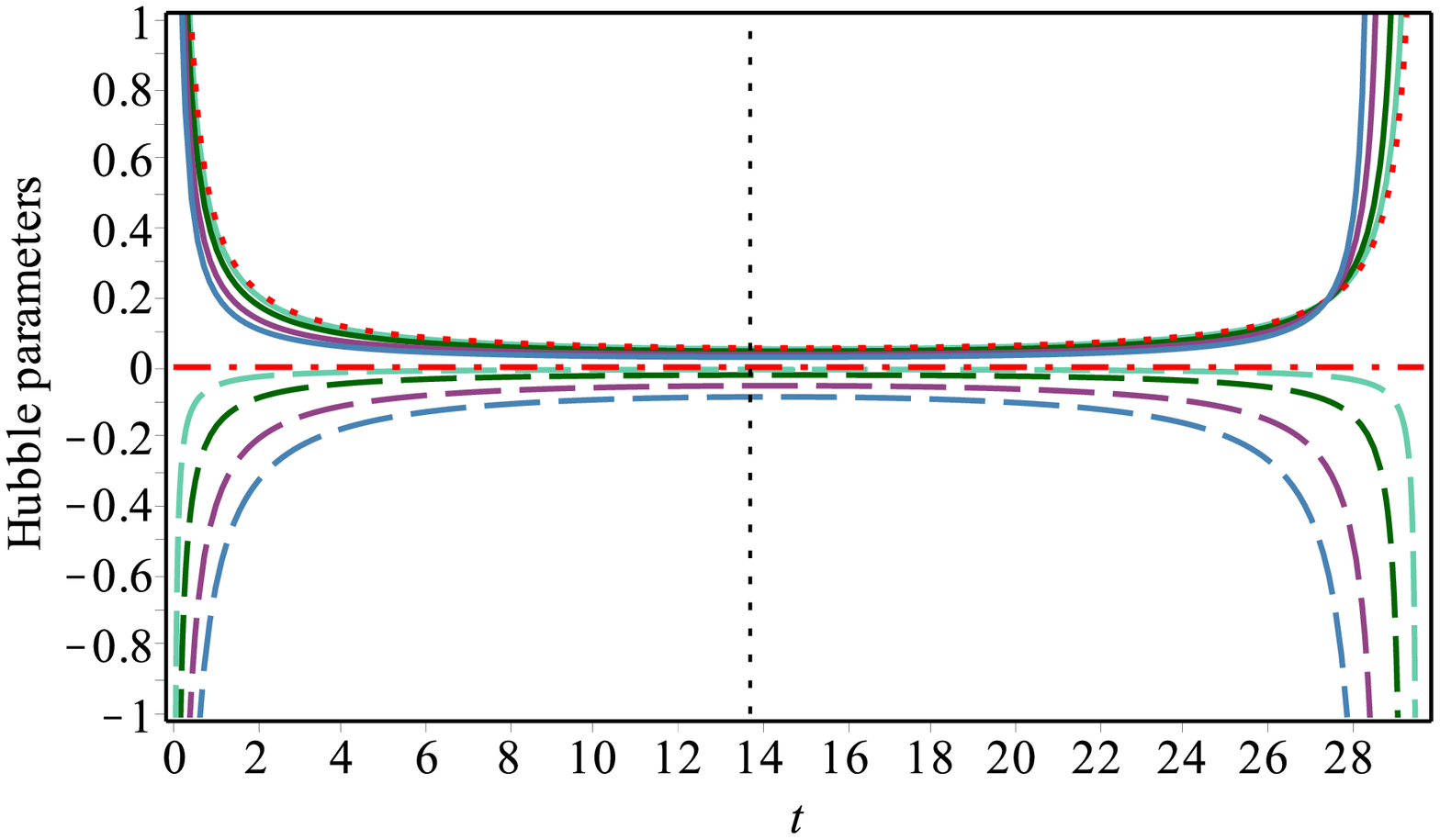}
\caption{Hubble parameters of the external and internal spaces versus cosmic time $t$ for different numbers of internal dimensions. We plotted for $n=1,2,6,22$ and for $n\rightarrow \infty$ limit, in the order from the solid blue curve to the dotted red curve.}
\label{fig:hw1b2}
\end{minipage}

\begin{minipage}[b]{0.49\linewidth}
\centering
\includegraphics[width=1\textwidth]{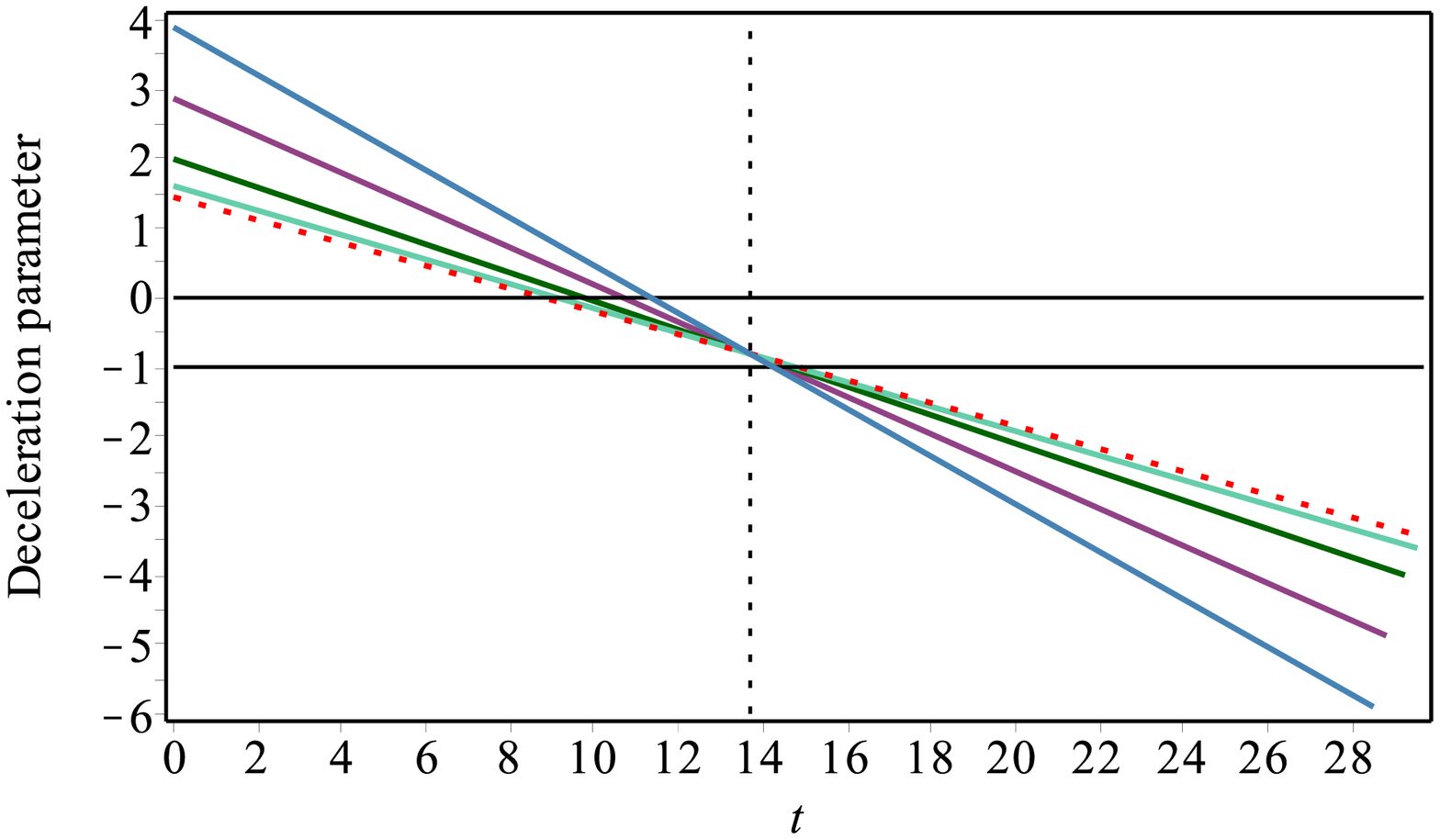}
\caption{Deceleration parameters of the external space versus cosmic time $t$ for different numbers of internal dimensions. We plotted for $n=1,2,6,22$ and for $n\rightarrow \infty$ limit, in the order from the solid blue curve to the dotted red curve.}
\label{fig:qw1b2}
\end{minipage}
\hspace{0.01\linewidth}
\begin{minipage}[b]{0.49\linewidth}
\centering
\includegraphics[width=1\textwidth]{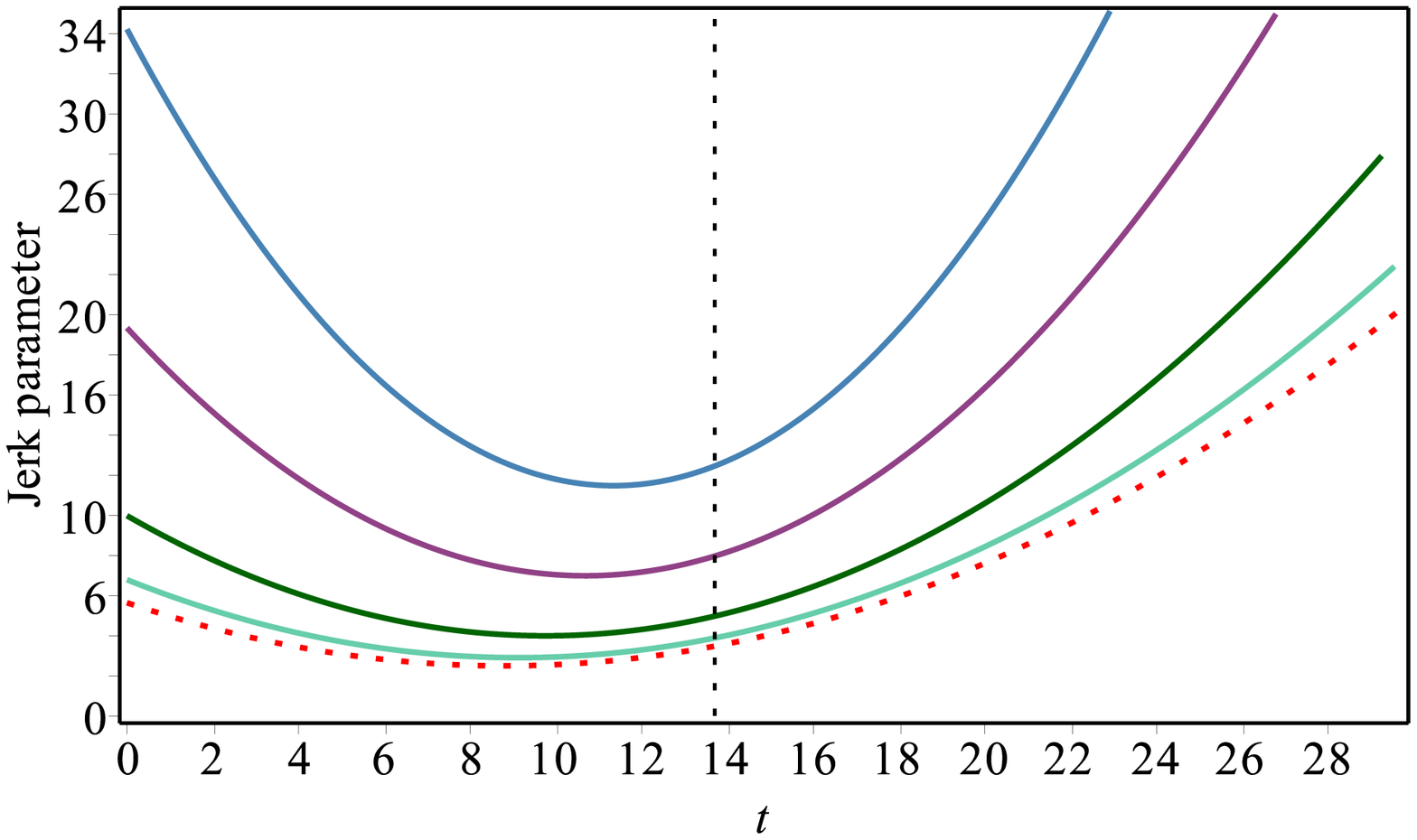}
\caption{Jerk parameters of the external space versus cosmic time $t$ for different numbers of internal dimensions. We plotted for $n=1,2,6,22$ and for $n\rightarrow \infty$ limit, in the order from the solid blue curve to the dotted red curve.}
\label{fig:jw1b2}
\end{minipage}
\end{figure}

\begin{figure}
\centering
\includegraphics[width=0.5\textwidth]{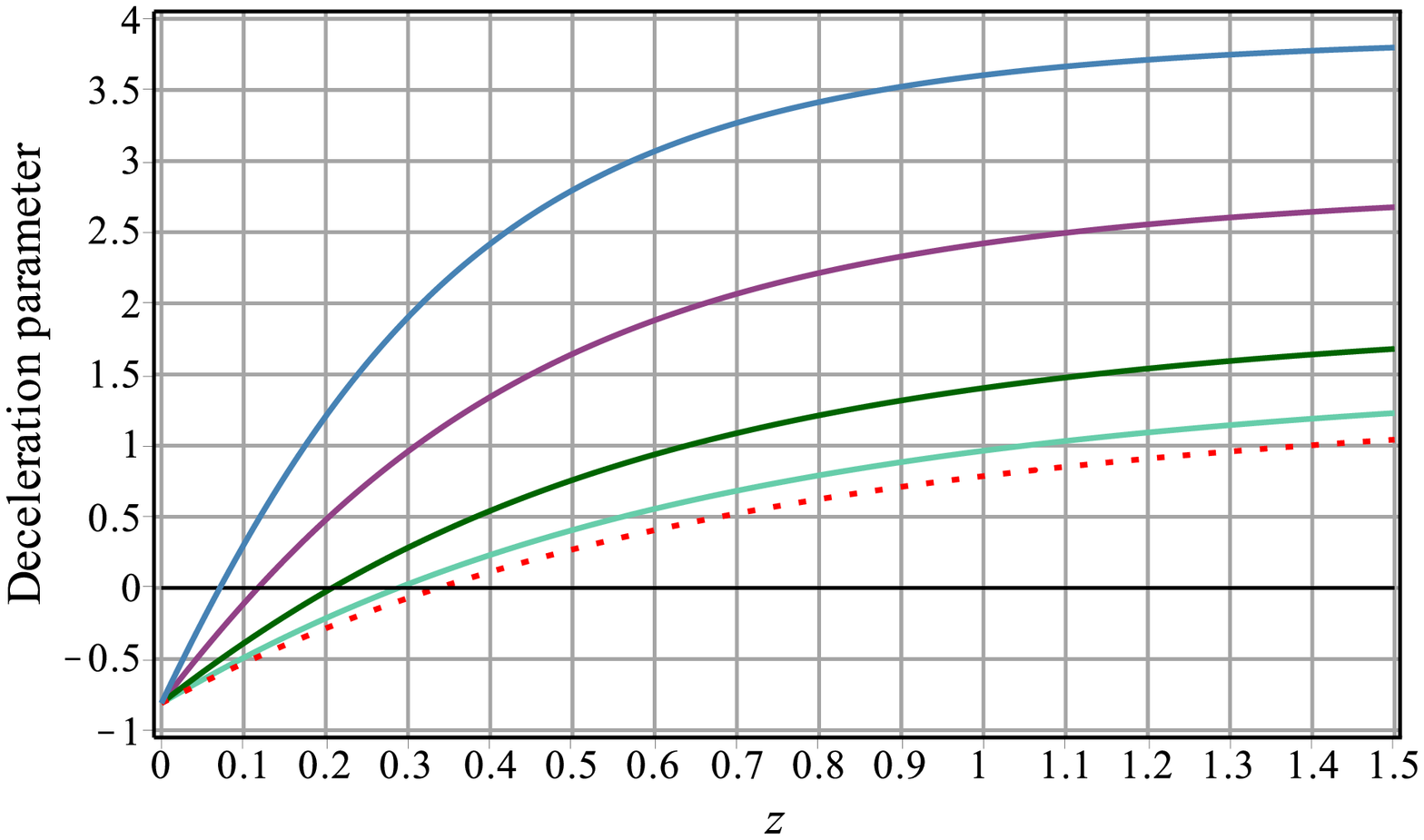}
\caption{Deceleration parameters of the external space versus cosmic redshift $z$ for different numbers of internal dimensions. We plotted for $n=1,2,6,22$ and for $n\rightarrow \infty$ limit, in the order from the solid blue curve to the dotted red curve.}
\label{fig:qw1zb2}
\end{figure}

The evolution of the universe starts with a zero size external space and infinitely large internal space at $t=0$. The external space expands until it becomes infinitely large at a finite time $t_{\rm{end}}=\frac{2\sqrt{2\mathcal{E}}}{U_{0}}$, while the internal space contracts to unobservable scales and then hits a singularity at $t_{\rm end}$. The external space enters into the accelerated expansion phase at $t_{\rm acc}=\frac{\sqrt{\mathcal{E}}}{U_{0}}\left(\sqrt{2}-\frac{1}{3}\sqrt{\frac{3n}{3+n}} \right)$ and into the super-exponential expansion phase at $t_{\rm{se}}=\frac{\sqrt{2\mathcal{E}}}{U_{0}}$. We note that all the cosmological parameters are determined by the values of $\frac{U_{0}}{\sqrt{\mathcal{E}}}$ and $n$ only; in particular $\frac{U_{0}}{\sqrt{\mathcal{E}}}$ determines the lifetime of the universe, $n$ determines the initial and final values of the deceleration parameters and they together determine the time of the onset of the acceleration of the external space. Hence, in principle, in this model also we are able to obtain the number of internal dimensions $n$ and $\frac{U_{0}}{\sqrt{\mathcal{E}}}$ upon obtaining the values of the deceleration parameter at the two different cosmic redshifts or two different times from observations. As in the previous section, choosing $q_{a,0}=-0.81$ and $t=13.7\,{\rm Gyr}$ considering the latest observations, we are able to depict the dynamics of the universe for different numbers of internal dimensions $n$. Accordingly, we plot the scale factors of the external and internal dimensions in Fig. \ref{fig:sfw1b2} and the Hubble parameters in Fig. \ref{fig:hw1b2} for $n=1,\;2,\;6,\;22$ and also for $n\rightarrow \infty$ limit. We plot the dimensionless deceleration parameter of the external dimensions in Fig. \ref{fig:qw1b2} and the jerk parameter in Fig. \ref{fig:jw1b2}. The evolution of the deceleration parameter of the external space with respect to the cosmic redshift $z$,
\begin{equation}
\label{eqn:QZ}
q_{a}(z) = -1+\sqrt{6}\sqrt{\frac{3+n}{n}}\tanh \left[\frac{\sqrt{6}}{2}\sqrt{\frac{3+n}{n}}\ln(z+1)+{\rm{arctanh}}\left(\frac{1}{\sqrt{6}}\sqrt{\frac{n}{3+n}}(1+q_{a}(z=0))\right)\right]
\end{equation}
is depicted in Fig. \ref{fig:qw1zb2}. We summarize the plots by giving the cosmic redshift ($z_{\rm acc}$) and the time passed since the time of the onset of acceleration ($13.7\,{\rm Gyr}-t_{\rm acc}$), the time of the Big Rip ($t_{\rm end}$) and the initial ($q_{a,{\rm int}}$, $j_{a,{\rm int}}$), present ($q_{a,0}$, $j_{a,0}$) and final ($q_{a,{\rm end}}$, $j_{a,{\rm end}}$) values of the deceleration and jerk parameters in Table \ref{table:2}.


\begin{table}[t]\footnotesize
  \caption{Cosmological parameters for the case $w=\frac{1}{2}$ ($\lambda=2$) for different numbers of internal dimensions $n$ assuming $q_{a,0}=-0.81$ at $t_{\rm today}=13.7\,{\rm Gyr}$}
  \label{table:2}
\begin{center}
\begin{tabular}{cc|c|c|c|c|c|c|c|c|l}
\cline{2-6}
\multicolumn{1}{l|}{} & \multicolumn{1}{c|}{$n=1$} & \multicolumn{1}{c|}{$n=2$}  & \multicolumn{1}{c|}{$n=6$} & \multicolumn{1}{c|}{$n=22$} & \multicolumn{1}{c|}{$n\rightarrow \infty$} \\[3pt]
\cline{2-6}
\cline{1-6}
\multicolumn{1}{|l|}{$z_{\rm acc}$ } & 0.07  & 0.12  & 0.21 & 0.29 & 0.34 \\[3pt]
\multicolumn{1}{|l|}{$13.7\,{\rm Gyr}-t_{\rm acc}$} & 2.36 Gyr & 3.01 Gyr & 3.95 Gyr & 4.58 Gyr  & 4.91 Gyr\\[3pt]
\multicolumn{1}{|l|}{$t_{\rm end}$} & 28.51 Gyr & 28.81 Gyr & 29.25 Gyr & 29.55 Gyr & 29.70 Gyr \\[3pt]
\cline{1-6}
\multicolumn{1}{|l|}{$q_{a,{\rm int}}$, $j_{a,{\rm int}}$} & 3.90, 34.30  & 2.87, 19.38 & 2.00, 10.00 & 1.61, 6.80 & 1.45, 6.65 \\[3pt]
\multicolumn{1}{|l|}{$q_{a,0}$, $j_{a,0}$} & -0.81, 12.48 & -0.81, 7.98 & -0.81, 4.98 & -0.81, 3.89 & -0.81,  3.48 \\[3pt]
\multicolumn{1}{|l|}{$q_{a,{\rm end}}$, $j_{a,{\rm end}}$} & -5.90, 63.70  & -4.87, 42.62 & -4.00, 28.00 & -3.61, 22.47 & -3.45, 20.35 \\[3pt]
\cline{1-6}
\multicolumn{1}{|l|}{$s_{0}$} & $<10^{-43}\,{\rm m}$ & $<10^{-30}\,{\rm m}$ & $<10^{-21}\,{\rm m}$ & $<10^{-18}\,{\rm m}$ & $<10^{-16}\,{\rm m}$ \\[3pt]
\multicolumn{1}{|l|}{$Y_{\rm p}$ ($^4$He mass fraction)} & $0.1538\pm 0.0006$   & $0.1711\pm 0.0006$ & $0.1952\pm 0.0006$ & $0.2111\pm 0.0006$ & $0.2191\pm 0.0006$ \\[3pt]
\multicolumn{1}{|l|}{$t_{T\sim 80\,{\rm keV}}$} & $10^{-24}\, {\rm s}<887\,{\rm s}$   & $10^{-16}\, {\rm s}<887\,{\rm s}$ & $10^{-8}\, {\rm s}<887\,{\rm s}$ & $10^{-5}\, {\rm s}<887\,{\rm s}$ & $10^{-3}\, {\rm s} <887\,{\rm s}$ \\[3pt]
\cline{1-6}
\end{tabular}
\end{center}
\end{table}

We observe that while the number of the internal dimensions alters the expansion rate of the external dimensions slightly, it alters the contraction rate of the internal dimensions drastically. We note that the deceleration parameter of the external space yields a form that is linear in time with a negative slope. This form of the deceleration parameter was suggested phenomenologically by Akarsu and Dereli \cite{AkarsuDereliLVDP1,AkarsuDereliLVDP2} to generalize the cosmological models with constant deceleration parameter so as to obtain kinematics that are consistent with the more recent observations. In that study, both the initial value and the slope of the deceleration parameter were arbitrary. On the other hand, here the initial value of the deceleration parameter is determined solely by the number of internal dimensions $n$, namely $q_{a,{\rm int}}=2\sqrt{6}-1,...,\sqrt{6}-1$ for $n=0,...,\infty$. From Table \ref{table:2}, one may observe that it is more likely that our model fits the observed dynamics of the universe when the number of the internal dimensions is high. Accordingly, the cases $n=1$ and $n=2$ can be excluded, since $q_{a,{\rm int}}>2$. One may check from Table \ref{table:2} that the transition to the accelerated expansion is indeed too recent and the jerk parameters evolves with very high values up to the present time of the universe. We would like to note an interesting property for the case of $n=6$ that corresponds to the number of spacetime dimensions in anomaly-free superstring theories. In this case the initial value of the deceleration parameter of the external dimensions gets exact integer values $q_{a,{\rm int}}=2$ and $j_{a,{\rm int}}=10$ which correspond to values of the deceleration parameter and jerk parameter in the presence of a stiff fluid in general relativity that was proposed by Zeldovich \cite{Zeldovich62} and then Barrow \cite{Barrow78} as the most probable EoS parameter at the very early universe. However, in this case the transition redshift is quite recent $z_{\rm acc}=0.21$ but it can still be accommodated within some observational studies (e.g. \cite{Alam04,Sahni06,Tasos09,Lima10}. In the case for $n=22$, both the transition redshift to the accelerated expansion $z_{\rm acc}=0.29$ and the value of the jerk parameter $j_{a,0}=3.86$ are in line with some observational studies \cite{Alam04,Sahni06,Tasos09,Lima10}.

In this solution, our models may accommodate the observed dynamics \cite{Alam04,Sahni06,Tasos09,Lima10,Cunha09,Li11,Visser04,Rapetti07,Cattoen08,Wang09,Vitagliano10,Capozziello11,Xia12} for the numbers of internal space larger than $n=6$. On the other hand, when we consider the early times of the universe, we observe that this solution for $w=\frac{1}{2}$ is unsucessful for all values of $n$, in contrast to the solution for $w=1$. It is possible to assure the universe to be effectively four dimensional during the BBN; one may see the maximum size of the internal dimensions today $s_{0}$ subject to the condition $s_{\rm NS}\lesssim l_{\rm proton}$ in Table \ref{table:2}. However, we cannot recover acceptable expansion rates for the 3-space in the early universe in accordance with a successful BBN. The deceleration parameter of the 3-space at the time of BBN can safely be taken as $q_{a=0}$ and hence the equation (\ref{eqn:qSteigman}) can be used for predicting the helium mass fraction. One may see from Table \ref{table:2}, however, that the predicted $Y_{\rm p}$ values accommodate neither the observationally inferred helium mass fraction $Y_{\rm p}=0.2565\pm 0.0060$ nor the helium mass fraction predicted by the SBBN $Y_{\rm p}=0.2485\pm 0.0006$, even within the error regions. One may also observe in Table \ref{table:2}, considering the SBBN, that the time $t_{T\sim 80\,{\rm keV }}$ is too early for any values of $n$.

\section{The universe according to an observer in $(1+3)$-dimensions}
\label{ObserverUniverse}
We argued above that the presence of internal dimensions should not affect the local physical processes, for instance the primordial nucleosynthesis, and hence cannot be observed directly and locally, for $n\gtrsim 2$ if they remain at Planck length scales today and for  $n\gtrsim 7$ if they remain at just below the scales probed by the LHC experiments today. Hence we can consider an observer who lives in a higher dimensional steady state universe but is unaware of the presence of the internal dimensions. Let us now assume that this observer considers the Einstein's general theory of relativity as the valid theory of gravity at cosmological scales. Because the observer cannot perceive the internal dimensions she/he would represent the geometry of the universe she/he perceives with the $(1+3)$-dimensional part of the metric (\ref{eqn:HDmetric}) we considered, i.e., with the spatially flat Robertson-Walker spacetime. Using the conventional general relativity by considering a comoving perfect fluid that would control the dynamics of the observed universe, the observer would obtain the following field equations:
\begin{equation}
\label{eqn:obsrhop}
3\frac{\dot{a}^2}{a^2}={\tilde{k}}\tilde{\rho}\quad\textnormal{and}\quad\frac{\dot{a}^2}{a^2}+2\frac{\ddot{a}}{a}=-{\tilde{k}}\tilde{p},
\end{equation}
where ${\tilde{k}}=8\pi \tilde{G}$ with $\tilde{G}$ being the gravitational constant scaled in 4-dimensions and $\tilde{\rho}$ and $\tilde{p}$ are the effective energy density and pressure inferred by the observer. In cosmology we can usually characterize a fluid with a simple equation of state parameter, which is the ratio between the pressure and energy density of the fluid and not necessarily constant. Hence, the observer is further able to obtain the equation of state parameter of the inferred effective fluid using the energy density and pressure:
\begin{equation}
\label{eqn:obsw}
\tilde{w}\equiv \frac{\tilde{p}}{\tilde{\rho}}=-\frac{1}{3}+\frac{2}{3}q_{a}.
\end{equation}
We apply this argument to the exact solutions we gave for $w=1$ and $w=\frac{1}{2}$.

\begin{figure}
\begin{minipage}[h]{0.49\linewidth}
\centering
\includegraphics[width=1\textwidth]{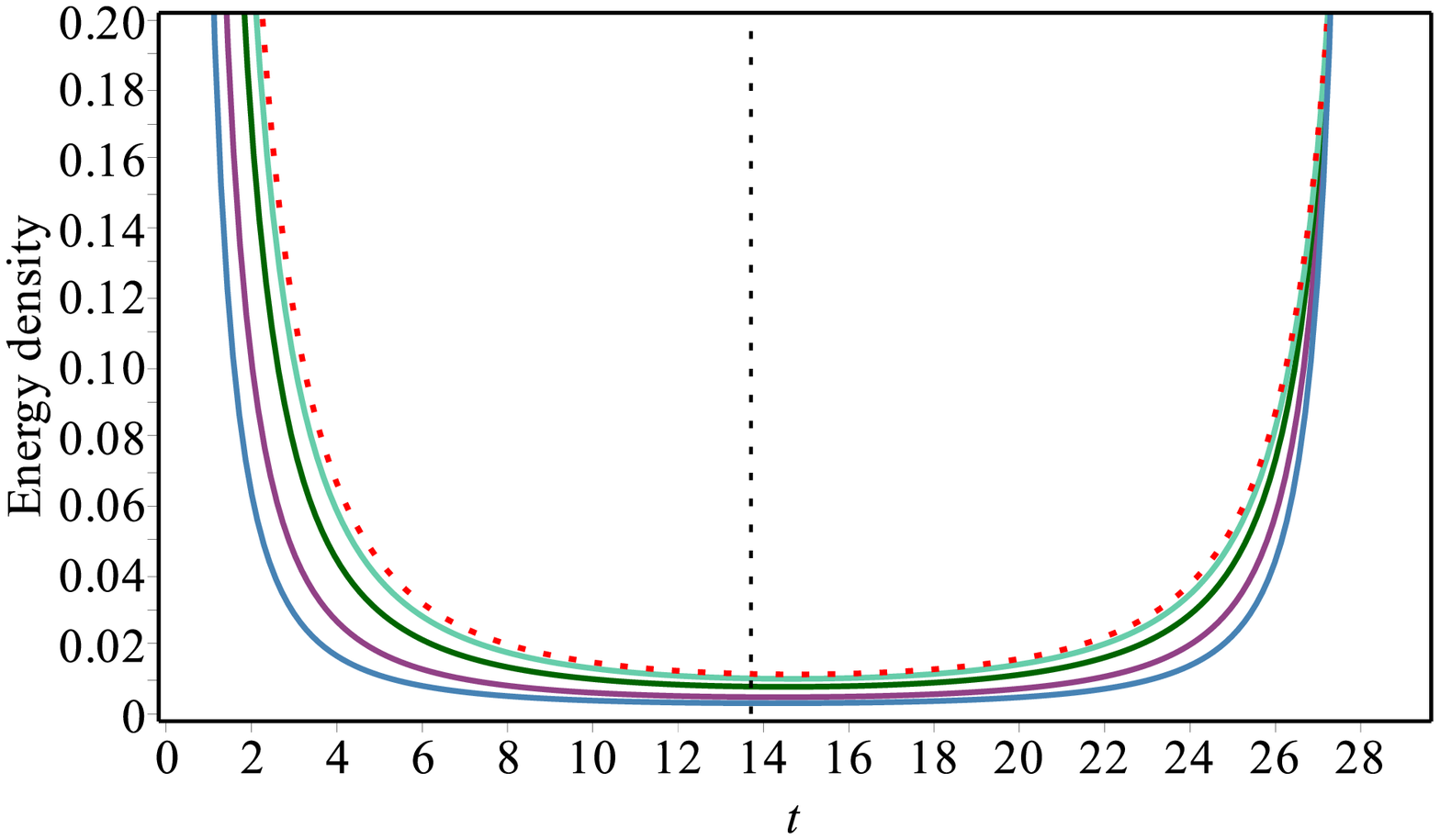}
\caption{The inferred energy density $\tilde{\rho}$ by an observer living in $(1+3)$-dimensions versus cosmic time $t$ for different numbers of internal dimensions. We plotted for $n=1,2,6,22$ and for $n\rightarrow \infty$ limit, in the order from the solid blue curve to the dotted red curve.}
\label{fig:obsrho1}
\end{minipage}
\hspace{0.01\linewidth}
\begin{minipage}[h]{0.49\linewidth}
\centering
\includegraphics[width=1\textwidth]{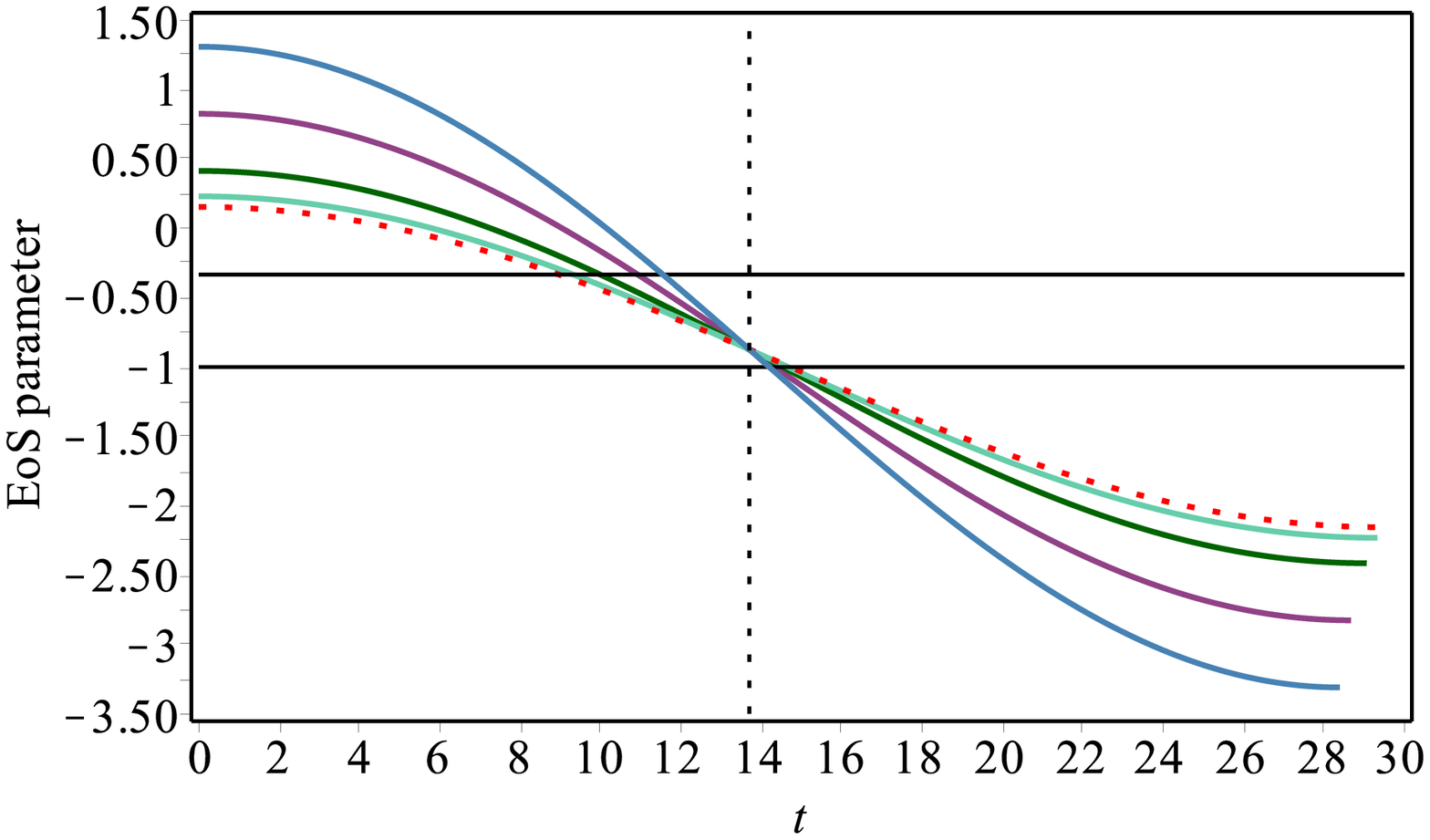}
\caption{The EoS parameter of the inferred fluid by an observer living in $(1+3)$-dimensions versus cosmic time $t$ for different numbers of internal dimensions. We plotted for $n=1,2,6,22$ and for $n\rightarrow \infty$ limit, in the order from the solid blue curve to the dotted red curve.}
\label{fig:obsw1}
\end{minipage}
\end{figure}

\bigskip 
\textbf{In the model with $\mathcal{E}>0$ and $w=1$ ($\lambda=0$)}, using $a$ we obtained for the case $w=1$ (\ref{eqn:scalesw1}) in (\ref{eqn:obsrhop}) and (\ref{eqn:obsw}), the energy density and EoS parameter of the induced four dimensional effective fluid become
\begin{equation}
\tilde{\rho}=\frac{1}{\tilde{k}}\frac{nU_{0}}{(3+n)\sin^{2}(\sqrt{U_{0}}\; t)}\quad\textnormal{and}\quad\tilde{w}=-1+2\sqrt{\frac{3+n}{3n}}\;\cos(\sqrt{U_{0}}\; t),
\end{equation}
respectively. The EoS above characterizes a fluid that would give the same kinematics as in general relativity for the external space as in our higher dimensional steady state model in dilaton gravity in the source-free universe. We plot the energy density of the inferred fluid in Fig. \ref{fig:obsrho1} and the EoS parameter in Fig. \ref{fig:obsw1} for different numbers of the internal dimensions $n$ by setting $q_{a,0}=-0.81$ at $t_{\rm today}=13.7\,{\rm Gyr}$ for the present age of the universe.
Using $q_{a, 0}=-0.81$ in (\ref{eqn:obsw}) we find that the present value of the EoS parameter of the inferred effective fluid is $\tilde{w}_{0}=-0.87$, which is consistent with the observational studies. However, we note that the fluid in fact exhibits a quintom behavior that can evolve below the phantom divide line $w=-1$; namely, $\tilde{w}\rightarrow -1+2\sqrt{\frac{3+n}{3n}}$ at $a=0$ and $\tilde{w}\rightarrow -1-2\sqrt{\frac{3+n}{3n}}$ as $a\rightarrow \infty$ (See Ref. \cite{Nesseris07,Cai} for crossing the phantom divide line and quintom cosmology). For instance, the EoS parameter evolves from $\tilde{w}_{a=0}\sim0.23$ to $\tilde{w}_{a=\infty}\sim-2.23$ for $n\sim22$ in which we are able to provide a complete picture of the universe with correct light element abundances and accommodate the observed dynamics of the universe.

\begin{figure}
\begin{minipage}[b]{0.49\linewidth}
\centering
\includegraphics[width=1\textwidth]{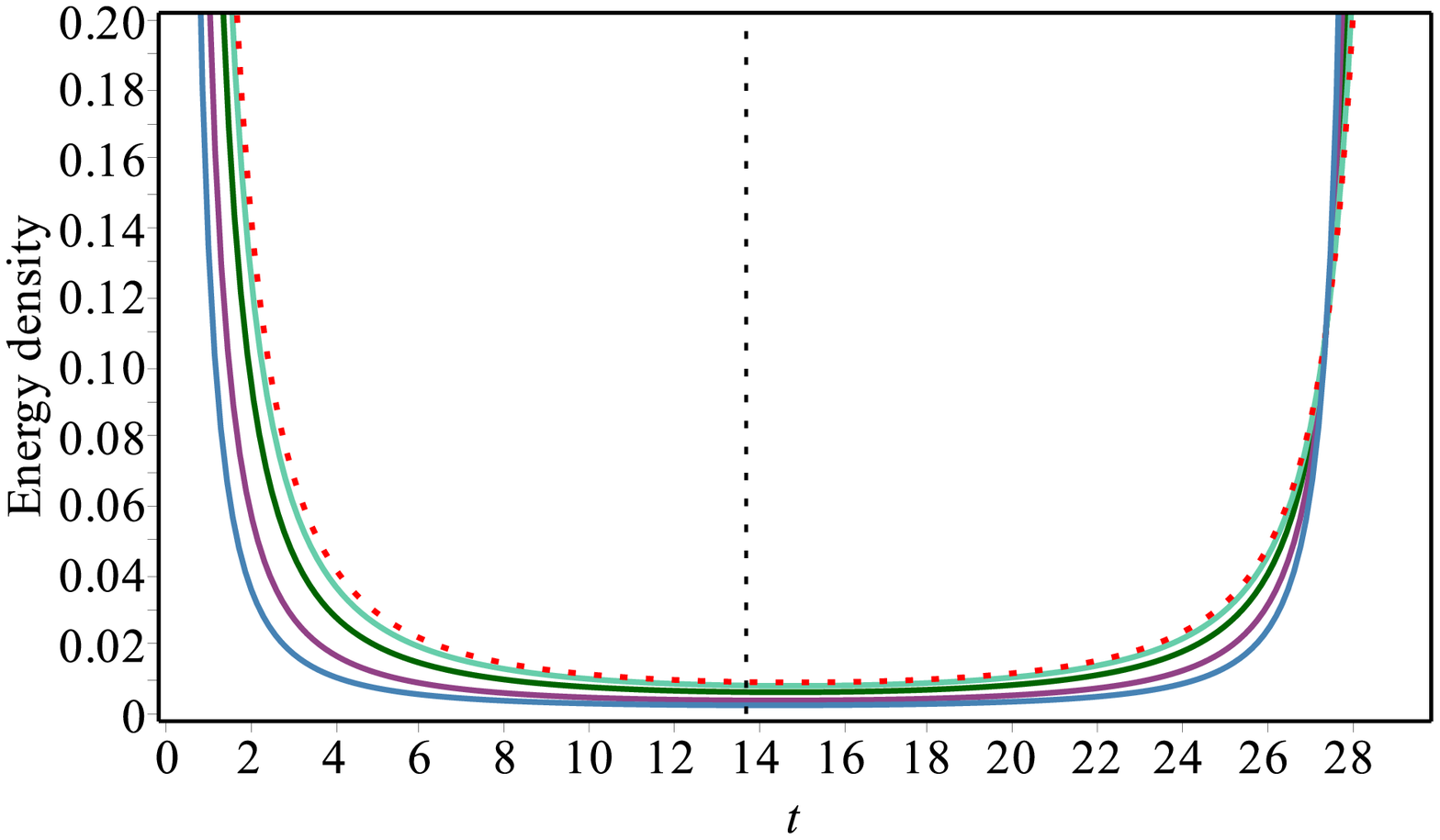}
\caption{The inferred energy density $\tilde{\rho}$ by an observer living in $(1+3)$-dimensions versus cosmic time $t$ for different numbers of internal dimensions. We plotted for $n=1,2,6,22$ and for $n\rightarrow \infty$ limit, in the order from the solid blue curve to the dotted red curve.}
\label{fig:obsrho1b2}
\end{minipage}
\hspace{0.01\linewidth}
\begin{minipage}[b]{0.49\linewidth}
\centering
\includegraphics[width=1\textwidth]{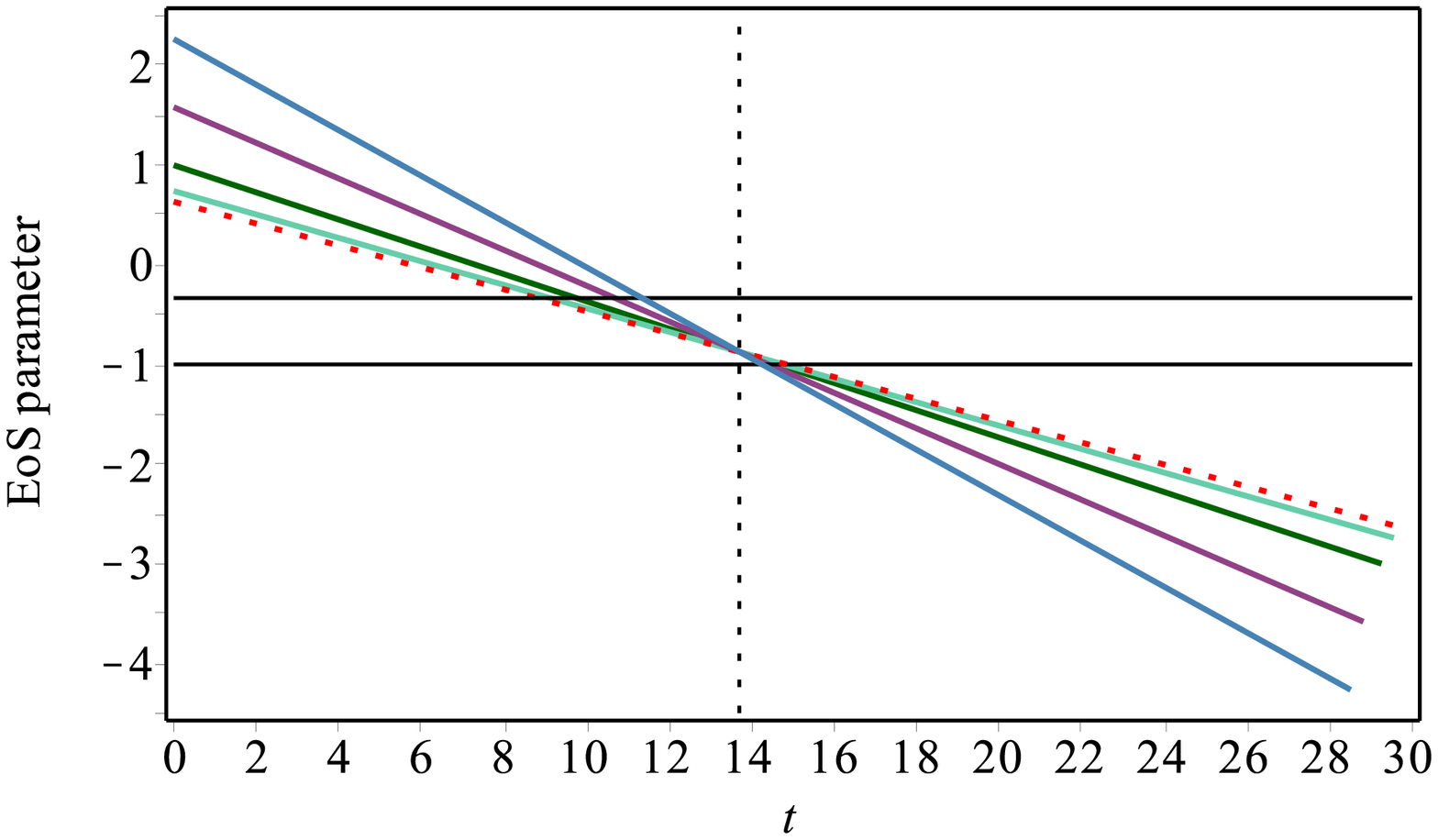}
\caption{The EoS parameter of the inferred fluid by an observer living in $(1+3)$-dimensions versus cosmic time $t$ for different numbers of internal dimensions. We plotted for $n=1,2,6,22$ and for $n\rightarrow \infty$ limit, in the order from the solid blue curve to the dotted red curve.}
\label{fig:obsw1b2}
\end{minipage}
\end{figure}

\bigskip 
\textbf{In the model with $\mathcal{E}>0$ and $w=\frac{1}{2}$ ($\lambda=2$)}, using $a$ (\ref{eqn:scalesw1b2}) we obtained for the case $w=\frac{1}{2}$ in (\ref{eqn:obsrhop}) and (\ref{eqn:obsw}), the energy density and EoS parameter of the inferred four dimensional effective fluid become
\begin{equation}
\tilde{\rho}=\frac{1}{\tilde{k}}\frac{4n\mathcal{E}}{(3+n)(-U_{0}\;t^2+2\sqrt{2\mathcal{E}}\;t)^{2}}
\quad\textnormal{and}\quad
\tilde{w}=-1-2\sqrt{\frac{3+n}{3n}}\frac{U_{0}}{\sqrt{\mathcal{E}}} \;t+2\sqrt{2}\sqrt{\frac{3+n}{3n}},
\end{equation}
respectively. The EoS here characterizes the fluid that would give the same kinematics as in general relativity for the external space in our higher dimensional steady state model in dilaton gravity in a source-free universe. We plot the energy density of the inferred fluid in Fig. \ref{fig:obsrho1b2} and the EoS parameter in Fig. \ref{fig:obsw1b2} for different numbers of the internal dimensions $n$ by setting $q_{a,0}=-0.81$ at $t_{\rm today}=13.7\,{\rm Gyr}$ for the present age of the universe.
Using $q_{a,0}=-0.81$ in (\ref{eqn:obsw}) we find that the present value of the EoS parameter of the inferred effective fluid is $\tilde{w}_{0}=-0.87$, which is consistent with the observational studies. In this solution we find the case $n=6$ interesting due to two reasons: (i) the initial value of the EoS parameter is $1$ that corresponds to a stiff fluid that was already proposed as the most probable EoS parameter of the very early universe \cite{Zeldovich62,Barrow78}, (ii) $1+3+n=10$ corresponds to gravi-dilaton action of the anomaly-free effective superstring theories.

\section{Discussion}
\label{Discussion}
Inspired by the low-energy effective string theory, we considered the source-free dilaton gravity action in $(1+3+n)$-dimensions in the string frame with a dilaton coupling constant $w>0$ that is arbitrary and a dilaton self-interaction potential that is in exponential form $U(\varphi) = U_0 e^{\lambda \varphi}$. In type IIB models the dilaton can be fixed dynamically and most phenomenological constructions based on these constructions assume a spatially constant dilaton. For instance it is possible to take the dilaton to be spatially constant but some of the K\"{a}hler moduli (corresponding to volumes) can evolve \cite{Biswas06}. Our model does not take into account e.g. fluxes, branes nor global effects, but the fact that these are technically possible in principle provides motivation of a phenomenological study of such a situation in which the total volume modulus is fixed, the volume moduli of the 6-cycle and 4-cycles are evolving, and the dilaton is spatially homogeneous. A model with the inflation driven by a K\"{a}hler modulus corresponding to the internal space volume which is shrinking, while the 3D volume is inflating is in fact studied \cite{Conlon06}.

We take our motivation from an interesting class of higher dimensional cosmological solutions that yield constant higher dimensional volume \cite{Freund82,DereliTucker83,BleyerZhuk96,RainerZhuk00,HoKephart10}, and solved the field equations by assuming that the higher dimensional volume scale factor of the universe is constant. This was the main assumption throughout the paper which is crucial (see \cite{AkarsuDereli12HDSS} for a futher discussion on this assumption). Here we study a phenomenological model aiming at exploring the possible outcomes of these constructions and we are not proposing a detailed fundamental model. However, our work is inspired by recent progress in string theory constructions, where dynamical mechanisms for stabilizing K\"{a}hler moduli have recently been developed \cite{Kachru03}. Thus even though our constant volume assumption may not be a generic or natural feature of such models, it appears to be technically possible to implement it within a fundamental theory. For example in type IIB theory, the KKLT construction allows to stabilize the K\"{a}hler moduli through non-perturbative effects on D-branes and in the simplest models one considers a single K\"{a}hler modulus which corresponds to the overall internal volume \cite{Conlon06} and in more complicated models, moduli corresponding to volumes of certain cycles are stabilized while moduli of other cycles can evolve dynamically \cite{Watson03}. We do not directly deal with these issues here, which would involve fluxes and global effects, but we simply assume that these could be arranged and explore what phenomenology could come out of this.

Under the above assumptions, we found that the system of the field equations is consistent provided $\lambda=4-4w$. A general explicit solution of the system could not be found but two solutions for arbitrary values of $w$ (namely, a solution where the space is static and another solution where the external space exhibits a power-law expansion) and two other solutions for $w=1$ and $w=\frac{1}{2}$ where the external space expands with a time varying deceleration parameter. We explored and constrained the parameter space of the dilaton coupling constant $w$ and the number of internal dimensions $n$ with a particular interest to determine which combination could better describe the expansion history of the observed universe. We showed that the internal dimensions could contract to such small sizes that the universe would appear effectively four dimensional starting from sufficiently early times, though they affect the evolution of the external space throughout the history of the universe. We focused on cases $w=1$ and $w=\frac{1}{2}$ and discussed these for various number of internal dimensions $n$, choosing the present age of the universe as $t_{0}=13.7$ Gyr and the present value of the deceleration parameter as $q_{a,0}=-0.81$ consistently with various observational studies. We further discussed whether the models give suitable dynamics for primordial nucleosynthesis processes to produce acceptable abundances of the light elements. The results obtained in the case $w=1$, which reduces the action to the gravi-dilaton effective string action, were more promising than the results obtained in the case $w=\frac{1}{2}$. The particular case $w=1$ and $1+3+n=26$ seems to be the most interesting one. In this case not only our action reduces to the gravi-dilaton action of the anomaly-free bosonic string theory, but also choosing  only the values $q_{a,0}=-0.81$ at $t_{0}=13.7$ Gyr from observations we found that the external space starts accelerating $4.38\,{\rm Gyr}$ ago from the present time at redshift value $z_{\rm acc}=0.31$, the present value of the jerk parameter will be $j_{a,0}=3.88$ consistently with observational studies. Moreover we found that the predicted  $^4$He mass fraction $Y_{\rm p}=0.2618\pm 0.0006$ is in good agreement with the inferred value $Y_{\rm p}=0.2565\pm 0.0060$ from the most useful observations for BBN \cite{Izotov10}, while the SBBN prediction is $Y_{\rm p}^{\rm SBBN}=0.2485\pm 0.0006$ \cite{Steigman08}. Considering the present temperature of the cosmic background radiation ($2.728\,{\rm K}$) we found that the energy scales of the primordial nucleosynthesis $\sim 80\,{\rm keV}$ is reached at $t\sim 94\,{\rm s}$, so that BBN model would work properly. We also have also a further interesting observation. The predicted value of $^4$He mass fraction is almost the same with the one that would be obtained when an extra neutrino type is included into the SBBN model. We finally comment on how the universe would be conceived by an observer in four dimensions who is unaware of the internal dimensions and assumes that the conventional general relativity is valid at cosmological scales.

\begin{figure}
\begin{minipage}[b]{0.49\linewidth}
\centering
\includegraphics[width=1\textwidth]{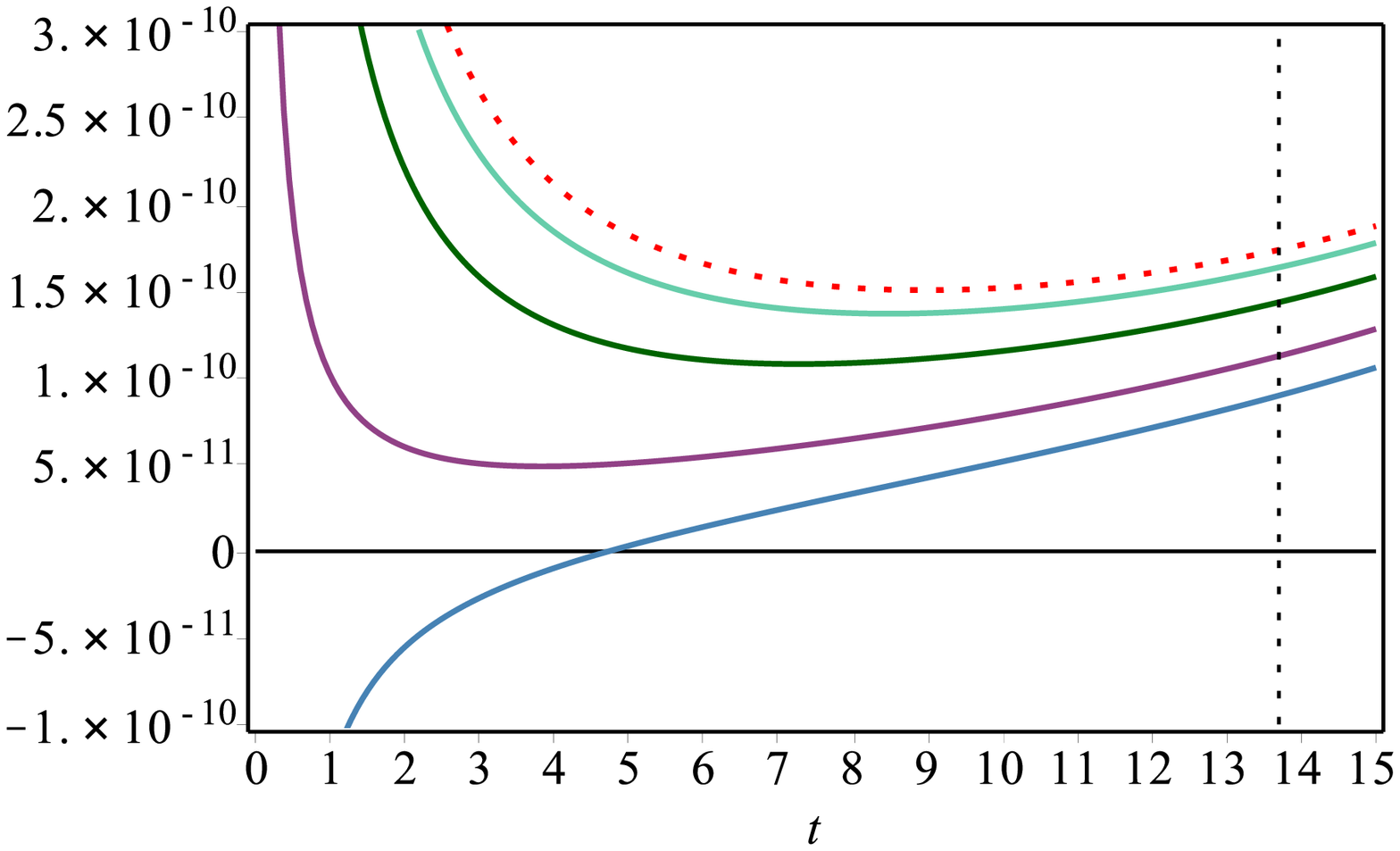}
\caption{The time variation of the $(1+3)$-dimensional gravitational constant versus cosmic time $t$ for different numbers of internal dimensions for $w=1$. We plotted for $n=1,2,6,22$ and for $n\rightarrow \infty$ limit, in the order from the solid blue curve to the dotted red curve.}
\label{fig:G4var1}
\end{minipage}
\hspace{0.01\linewidth}
\begin{minipage}[b]{0.49\linewidth}
\centering
\includegraphics[width=1\textwidth]{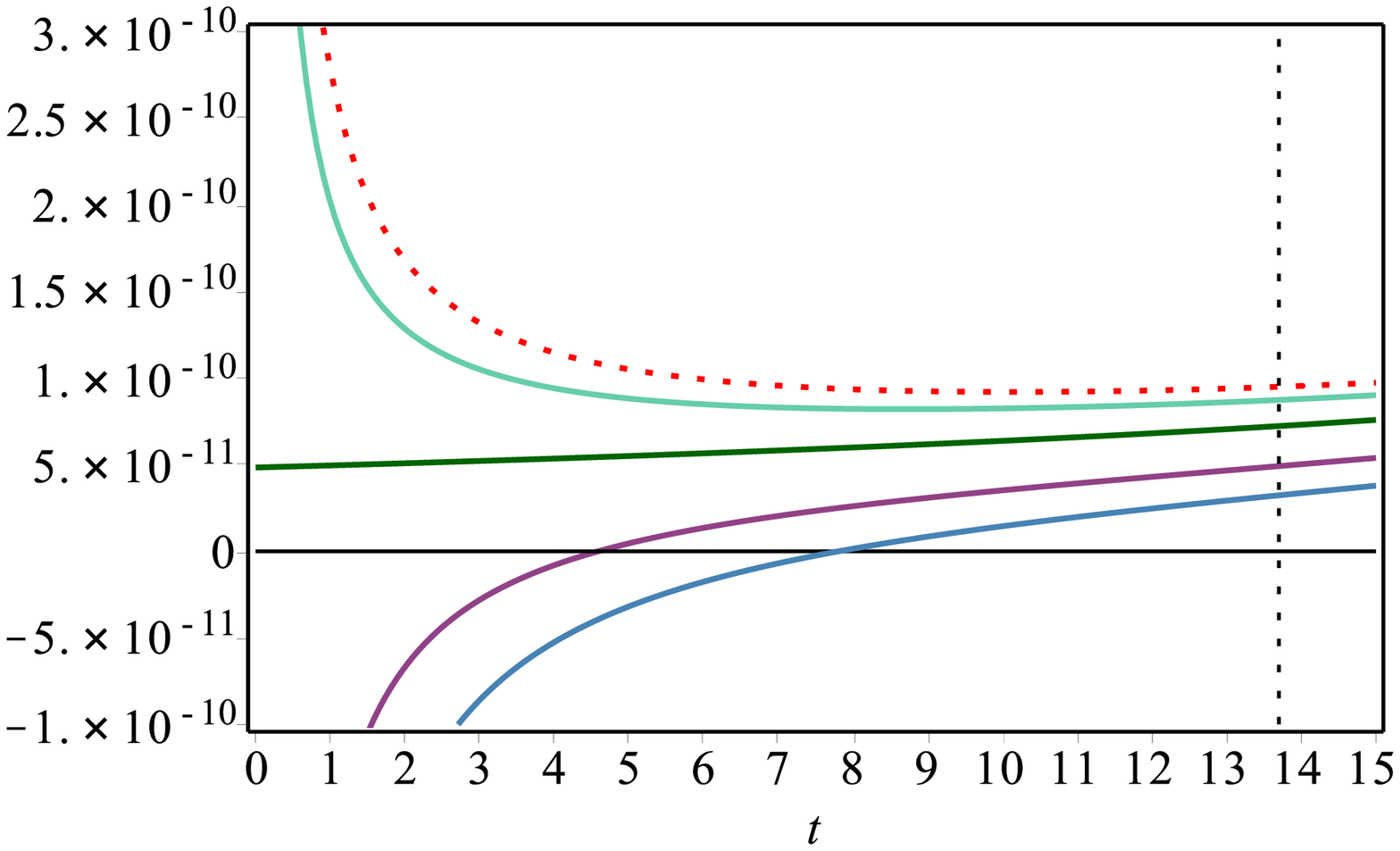}
\caption{The time variation of the $(1+3)$-dimensional gravitational constant versus cosmic time $t$ for different numbers of internal dimensions for $w=\frac{1}{2}$. We plotted for $n=1,2,6,22$ and for $n\rightarrow \infty$ limit, in the order from the solid blue curve to the dotted red curve.}
\label{fig:G4var1b2}
\end{minipage}
\end{figure}

The strength of the effective $(1+3)$-dimensional gravitational coupling depends both on the dilaton field and the volume scale factor of the compact internal space \cite{Gasperini07,Uzan11}. Hence, before concluding the paper, we wish to discuss the time variations of the $(1+3+n)$-dimensional gravitational coupling ($G_{1+3+n}$) and of the $(1+3)$-dimensional effective gravitational coupling ($G_{1+3}$) in our solutions given in Section \ref{sec:w1} and Section \ref{sec:w1b2}. The dilaton field in the action (\ref{eqn:action}) stands as a reciprocal of the $(1+3+n)$-dimensional gravitational coupling, that is,
\begin{equation}
e^{-2\varphi}=\beta \propto \frac{1}{G_{1+3+n}},
\end{equation}
which is obviously time dependent in all our solutions in which the universe is dynamical. Accordingly, time variation of the $(1+3+n)$-dimensional gravitational coupling will be
\begin{equation}
\label{eqn:GVdilaton}
\frac{G'_{1+3+n}}{G_{1+3+n}}=-\frac{\beta'}{\beta}=-(1+q_{a})H_{a}.
\end{equation}
Thus we find
\begin{equation}
\frac{G'_{1+3+n}}{G_{1+3+n}}=-\sqrt{U_{0}}\cot{(\sqrt{U_{0}}\,t)}\quad\textnormal{for } w=1
\end{equation}
and
\begin{equation}
\frac{G'_{1+3+n}}{G_{1+3+n}}=\frac{2U_{0}\,t-2\sqrt{2\mathcal{E}}}{-U_{0}\,t^2+2\sqrt{2\mathcal{E}}\,t}\quad\textnormal{for } w=\frac{1}{2}.
\end{equation}
We observe that the time variation of the $G_{1+3+n}$ is independent of the number of internal dimensions and is negative for $t< t_{\rm se}$, almost null (the action we considered mimics the Einstein-Hilbert action) when $t\sim t_{\rm se}$ (i.e. when $\beta\sim\beta_{\rm max}$) and positive for $t> t_{\rm se}$. Using $q_{a,0}=-0.81$ and $t_{0}=13.7\,{\rm Gyr}$, we found that its present value will be $\sim -10^{-11}\,{\rm yr}^{-1}$ in both cases $w=1$ and $w=\frac{1}{2}$. We note that our action deviates from the Einstein-Hilbert action considerably both at the very early and very late times of the universe. On the other hand, what concerns us here is the effective $(1+3)$-dimensional gravitational coupling rather than the $(1+3+n)$-dimensional gravitational coupling since the observed universe is effectively four dimensional. The geometry of the space has a Kaluza-Klein structure so that $G_{1+3}$ will be related to $G_{1+3+n}$ through the proper volume of the internal space as follows \cite{Gasperini07,Uzan11}:
\begin{equation}
\label{eqn:GVKaluza}
G_{1+3} \propto \frac{G_{1+3+n}}{V_{n}}.
\end{equation}
Hence, considering (\ref{eqn:GVdilaton}) and (\ref{eqn:GVKaluza}) together we find the time variation of the $(1+3)$-dimensional gravitational coupling is given by
\begin{equation}
\label{eqn:GVdilatonKaluza}
\frac{G'_{1+3}}{G_{1+3}}=-\frac{\beta'}{\beta}-n H_{s}=(2-q_{a})H_{a}.
\end{equation}
This implies
\begin{equation}
\frac{G'_{1+3}}{G_{1+3}}=-\sqrt{U_{0}}\left(\cot{(\sqrt{U_{0}}\,t)}-\sqrt{\frac{3n}{3+n}}\frac{1}{\sin{(\sqrt{U_{0}}\,t)}}\right) \quad\textnormal{for } w=1
\end{equation}
and
\begin{equation}
\frac{G'_{1+3}}{G_{1+3}}=\frac{2U_{0}\,t-2\sqrt{\mathcal{E}}\left(\sqrt{2}-\sqrt{\frac{3n}{3+n}}\right)}{-U_{0}\,t^2+2\sqrt{2\mathcal{E}}\,t}\quad\textnormal{for } w=\frac{1}{2}.
\end{equation}
We observe that the number of the internal dimensions affects the time variation of $G_{1+3}$ and can diversify its evolution considerably depending on the value of $n$. The time variation of $G_{1+3}$ is always positive for $n \geq 2$, but takes negative values at earlier times for $n=1$ in the case $w=1$; while it is always positive for $n\geq 6$, but takes negative values at earlier times for $n\leq 5$ in the case $w=\frac{1}{2}$. The case $w=\frac{1}{2}$ with $n=6$ is unique since it does not diverge as $t\rightarrow 0$. We depicted the time variation of $G_{1+3}$ per year for times between $t=0$ and $t=15\,{\rm Gyr}$ for $w=1$ in Fig. \ref{fig:G4var1} and for $w=\frac{1}{2}$ in Fig. \ref{fig:G4var1b2} by considering $q_{a,0}=-0.81$ and $t_{0}=13.7\,{\rm Gyr}$.
We observe that the time variation of $G_{1+3}$ today will be $\sim 1\times 10^{-10}\,{\rm yr}^{-1}$ in the case $w=1$ and $\sim 5\times 10^{-11}\,{\rm yr}^{-1}$ in the case $w=\frac{1}{2}$. It won't be higher than $\sim 10^{-10}\,{\rm yr}^{-1}$ in both cases when the universe was older than $\sim 1\,{\rm Gyr}$, which is the age of the universe when the first bound structures were formed.

Majority of the constraints coming from the Solar system, pulsar timing or stellar observations on the time variation of $G_{1+3}$ that is found in literature \cite{Uzan11} favor a value $\sim \pm10^{-11}\,{\rm yr}^{-1}$ in the vicinity of the present age of the universe. Considering the random and the systematic errors involved in the determination of the actual constraints, our results are still marginally consistent with the above value. On the other hand, the magnitude of the time variation of $G_{1+3}$, in our models, increases (except the case $w=\frac{1}{2}$ and $n=6$) and exceeds the observational limits considerably for times less than $\sim 1\,{\rm Gyr}$. However, the constraints on the variation of $G_{1+3}$ for times less than $\sim 1\,{\rm Gyr}$ were set from cosmological observations such as the primordial nucleosynthesis and/or CMB angular-power spectrum and are not applicable to our models. The reason being that, the time-dependence of $G_{1+3}$ does not affect directly the primordial nucleosynthesis and CMB angular-power spectrum but only indirectly through its possible effects on the expansion rate/history of the universe. The main idea behind setting constraints on the time variation of $G_{1+3}$ using these observations is as follows: firstly the constraints on the kinematics of the universe are set (for instance, by considering the requisite kinematics for a successful primordial nucleosynthesis proccess) and then the time variation of $G_{1+3}$ is constrained such that the kinematics is altered without violating the kinematical constraints considering a particular theory and/or cosmological model by considering the possible variation of $G_{1+3}$ (see for instance \cite{Barrow78b}). On the other hand, in our study, we have already explored our parameter space of $w$ and $n$ by considering whether the expansion of the external space could describe the expansion history of the universe successfully. Namely, using the present value of the deceleration parameter and the present age of the universe, we discussed whether the expansion of the external space can accommodate the observed kinematics of the recent universe. Then, because the $^4$He mass fraction is very sensitive to the expansion rate of the early universe, we calculated the predicted $^4$He mass fractions by considering the non-standard expansion rates of the early universe in our models and compared them with the observationally inferred $^4$He mass fraction values. We also discussed whether $T\sim 80\,{\rm keV}$ was reached before the free neutrons decay $t\lesssim 887\,{\rm s}$ to assure BBN model to work properly by considering the present CBR temperature ($2.728\,{\rm K}$), age ($13.7\,{\rm Gyr}$) and deceleration parameter value ($q_{a,0}=-0.81$) of the universe. We summarized our results for various values of the number of internal dimensions for the case $w=1$ in Table \ref{table:1} and for the case $w=\frac{1}{2}$ in Table \ref{table:2}. We observe that the case $w=1$ is superior to the case $w=\frac{1}{2}$ at describing both the present epoch and the past primordial nucleosynthesis epoch of the universe.

\begin{table}[t]\centering\footnotesize
\caption{The time-scales of some important energy-scales are calculated in the case $w=1$ for $n=1$ (the possible lowest number of internal dimensions), $n=6$ (the number of the internal dimensions in anomaly-free super-string theories), $n=22$ (the number of internal dimensions in anomaly-free bosonic string theory) and for $n\rightarrow\infty$ limit choosing the present time and deceleration parameter of the universe as $q_{\rm today}=-0.81$ and $t_{\rm today}=13.7\,{\rm Gyr}$. We consider the corresponding energy- and time-scales of some key events in the standard model for the history of the universe listed by Liddle and Lyth \cite{Liddle00}) and compare with our calculated values.}
\begin{tabular}{lllcccc}
\hline\hline
& Standard & Standard  & $w=1$ & $w=1$ & $w=1$ & $w=1$\\
Event & Energy-scale &  Time-scale & $n=1$ & $n=6$ & $n=22$ & $n\rightarrow\infty$
\\\hline\hline\\
First bound structures form. & $10^{-3}$ eV & $\sim 1$ Gyr & 0.1 Gyr & 1 Gyr & 1 Gyr & 1 Gyr\\[6pt]
Atom formation, photon decoupling (CMB). & $10^{-1}$ eV & $10^{5}$ yr &  $10$ yr & $10^{5}$ yr & $10^{5}$ yr & $10^{5}$ yr \\[6pt]
Matter-radiation equality. & 1 eV & $10^{4}$ yr & $10^{-2}$ yr & $10^{3}$ yr & $10^{4}$ yr & $10^{4}$ yr \\[6pt]
Primordial nucleosynthesis. & $10^{-1}$ MeV & $10^{2}$ s & $10^{-12}$ s & $10^{-1}$ s & $10^{2}$ s & $10^{2}$ s \\[6pt]
$\nu$ decoupling, $e\bar{e}$ annihilation. & 1 MeV & $1$ s & $10^{-16}$ s & $10^{-3}$ s & $1$ s & $1$ s \\[6pt]
$\gamma$, $\nu$, $e$, $\bar{e}$, $n$ and $p$ thermal equilibrium. & $10$ MeV & $10^{-2}$ s & $10^{-19}$ s & $10^{-5}$ s & $10^{-2}$ s & $10^{-1}$ s \\[6pt]
Quark-hadron phase transition? & $10^{2}$ MeV & $10^{-4}$ s &  $10^{-23}$ s & $10^{-7}$ s & $10^{-4}$ s & $10^{-2}$ s \\[6pt]
Electroweak phase transition? & $10^{2}$ GeV & $10^{-10}$ s  & $10^{-33}$ s & $10^{-13}$ s & $10^{-9}$ s & $10^{-8}$ s \\[6pt]
\hline\hline
\end{tabular}
\label{tab:keyevents}
\end{table}

We wish to conclude the paper by comparing the time evolution of the energy scales in the effective $(1+3)$-dimensional universe for $w=1$ with the one in the standard model for the history of the universe. We find the opportunity to calculate the time scale when a certain energy scale would be reached using (\ref{eqn:NS2}) and (\ref{eqn:scalesw1}). In our calculations we considered the energy-scales of some key events and the corresponding time-scales in the standard model for the history of the universe that are listed by Liddle and Lyth \cite{Liddle00}. We calculate the time-scale of the universe when the energy-scale of a specific key event given in the list considering the case $w=1$ for $n=1$ (the lowest possible number of internal dimensions), for $n=6$ (the number of internal dimensions in anomaly-free superstring theories), for $n=22$ (the number of internal dimensions in anomaly-free bosonic string theory) and as $n\rightarrow\infty$ choosing the present time and deceleration parameter of the universe as $t_{\rm today}=13.7\,{\rm Gyr}$ and $q_{\rm today}=-0.81$ consistent with observational studies. None of the time scales we obtain agrees with the time scales of standard cosmology for the case $n=1$. The time scales we obtain for $n=6$ agree with the time scales of the standard cosmology up to the energy scales $10^{-1}$ eV. In line with our previous discussions, we further note that the time scales we obtain for $n=22$ agree with the times scales of standard cosmology up to the energy scales $10^{2}$ GeV, i.e., up to time scales $\sim 10^{-10}\,{\rm s}$. In fact this is precisely the case for which our action coincides with the gravi-dilaton action of the anomaly-free bosonic string theory. It is also interesting to observe that as $n\rightarrow \infty$ the good agreement between the time scales fails. In that limit the obtained time scales agree with the time scales of standard cosmology only up to the energy scales $10^{-1}\,{\rm MeV}$.

It is remarkable that using only three observational values concerning the present day universe (the present age, value of the deceleration parameter and cosmic background radition temperature of the universe), among all possible values in the parameter space of the dilaton coupling $w$ and the number of the internal dimensions $n$, the case $w=1$ with $n\sim 22$ gives the best history for the effectively four dimensional universe that fits the standard model of the history of the universe \cite{Liddle00}. Moreover, the case $w=1$ and $n=22$ not only predicts the time and redshift when the accelerated expansion has started in accordance with the observations, but also predicts the $^4$He mass fraction in agreement with the mostly used observations for the Big Bang Nucleosynthesis \cite{Izotov10,Aver10}. All these provide a strong support for our crucial assumption throughout the paper that the higher dimensional space has a constant volume. We again emphasize that we do not introduce any higher-dimensional matter source in our action, which is in line with the central idea of string models where matter in four-dimensions is nothing but a manifestation of strings that live in higher dimensions. We are thinking of improving our model for instance by considering other string-motivated fields such as the axion and gauge fields (e.g. \cite{Sonner06,Witten06}) and ask the question whether constant higher dimensional volume could be related with a fundamental theory of physics.

\begin{center}
\textbf{Acknowledgments}
\end{center}
This research is supported by a Post-Doc Research Grant by the Turkish Academy of Sciences (T{\"{U}}BA). \"{O}.A. also acknowledges the support from Ko\c{c} University, and the financial support he received from, and hospitality of the Abdus Salam International Center for Theoretical Physics (ICTP), where part of this work is carried out. We thank Anastasios Avgoustidis, Anshuman Maharana and Fernando Quevedo for valuable discussions. We also thank the referee for criticisms that helped us to improve the paper.

\end{document}